\pdfoutput=1
\documentclass[11pt]{article}
\usepackage[preprint]{acl}

\usepackage{times}
\usepackage{latexsym}

\usepackage[T1]{fontenc}
\usepackage[utf8]{inputenc}
\usepackage{microtype}
\usepackage{inconsolata}
\usepackage{graphicx}
\usepackage{multirow}
\usepackage{booktabs}
\usepackage{array} 
\usepackage{makecell}
\usepackage{amsfonts}
\usepackage{amsmath}
\usepackage{amssymb}
\usepackage{bbm}
\usepackage{algorithm}
\usepackage{algorithmic}
\usepackage{pifont}

\newcommand{\cmark}{\ding{51}}
\newcommand{\xmark}{\ding{55}}

\usepackage[most]{tcolorbox}
\usepackage{caption}
\usepackage[table]{xcolor}
\usepackage{listings}

\definecolor{promptblue}{RGB}{0, 60, 255}
\definecolor{promptred}{RGB}{220, 0, 0}
\definecolor{qwenAudioBg}{RGB}{204,234,255}  
\definecolor{phiBg}{RGB}{245,238,255}
\definecolor{qwenOmniBg}{RGB}{232,247,242}
\definecolor{gemma3nBg}{RGB}{255,246,224}
\definecolor{gemma4Bg}{RGB}{255,235,220}
\definecolor{qwen3Bg}{RGB}{226,220,255}      
\definecolor{stepBg}{RGB}{238,245,235}
\definecolor{geminiBg}{RGB}{255,232,238}  

\newtcolorbox{promptbox}{
  enhanced,
  colback=white,
  colframe=black,
  boxrule=0.8pt,
  arc=0pt,
  outer arc=0pt,
  left=8pt,
  right=8pt,
  top=8pt,
  bottom=8pt,
  width=\linewidth
}

\newif\ifcomments
\commentstrue  

\newcommand{\gain}[1]{\textsubscript{\textcolor{green!60!black}{+#1}}}
\newcommand{\loss}[1]{\textsubscript{\textcolor{red!70!black}{#1}}}
\newcommand{\best}[1]{\textbf{#1}}
\newcommand{\second}[1]{\underline{#1}}

\title{\textsc{AudioProcessBench}: Benchmark for Identifying Process Errors in Audio-Grounded Reasoning}

\author{
 \textbf{Xiangyu Zhao\textsuperscript{1}},
 \textbf{Junyu Yan\textsuperscript{1}},
 \textbf{Yaling Shen\textsuperscript{1}},
 \textbf{Zimu Wang\textsuperscript{2}},
 \textbf{Yiwen Jiang\textsuperscript{1}},
 \textbf{Stephanie Fong\textsuperscript{1}},
\\
 \textbf{Qingyang Xu\textsuperscript{1}},
 \textbf{Jiahe Liu\textsuperscript{1}},
 \textbf{Dominic Dwyer\textsuperscript{1,3,4}},
 \textbf{Zongyuan Ge\textsuperscript{1}}
\\
\\
 \textsuperscript{1}Monash University,
 \textsuperscript{2}Xi'an Jiaotong-Liverpool University,
 \\
 \textsuperscript{3}Orygen,
 \textsuperscript{4}University of Melbourne
\\
}

\begin{document}
\maketitle
\begin{abstract}
Large audio-language models (LALMs) increasingly use explicit reasoning traces for complex audio understanding, yet the evaluation of reasoning quality remains underexplored. Although process-level benchmarks for process reward models (PRMs) have advanced reasoning evaluation in text and multi-modal domains, comparable evaluation for audio reasoning remains limited. In this paper, we present \textsc{AudioProcessBench}, a comprehensive benchmark for step-level process error identification in audio reasoning. \textsc{AudioProcessBench} contains diverse reasoning traces generated by 6 audio and omni language models. Each trace is segmented into discrete reasoning steps and annotated with binary step correctness and fine-grained error types. Our benchmark evaluates models under three complementary paradigms: (1) step correctness identification, (2) error-type-conditioned detection for diagnosing audio-specific verifier capacities, and (3) chain-level aggregation, where verifiers select or aggregate among multiple reasoning traces for the same question. This design enables a systematic analysis of whether current models can detect process errors, whether their weaknesses differ across audio-specific error types, and whether process verification translates into improved answer selection. \textsc{AudioProcessBench} provides a testbed for future research on audio reasoning verifiers, process reward models, and reliable omni-modal reasoning.
\end{abstract}

\section{Introduction}
Large language models (LLMs) have made remarkable progress on complex reasoning tasks such as mathematics, programming, and planning \citep{hadi2023large, raiaan2024review, luo2025large}. Multi-modal large language models (MLLMs) extend these reasoning capabilities to real-world signals, including images, audio, and video \citep{li2025systematic}. As reasoning models increasingly rely on explicit reasoning traces and test-time scaling, process-level verification has become crucial for assessing and improving model reasoning \citep{xu2025toward}. Modern reasoning systems therefore commonly use verifiers such as rule-based checkers, learned reward models, and LLM-as-critic models to evaluate intermediate reasoning steps \citep{wen2025reinforcement, yu2025reward}.

While such verification mechanisms have been extensively studied in text and vision-language reasoning, their effectiveness in audio-grounded reasoning remains underexplored. Existing audio reasoning benchmarks mostly evaluate final-answer accuracy, providing limited insight into whether a model's intermediate reasoning is faithful to the acoustic evidence \citep{su2025audio}. Conversely, recent process-level benchmarks for text and vision-language reasoning evaluate step correctness, reward-model behavior, or multi-modal reasoning errors \citep{zheng2025processbench, song2025prmbench, zhou2025mpbench, wang2025visualprm, jacovi2024chain}. However, these benchmarks do not address the distinctive failure modes of audio-grounded reasoning, such as hallucinated acoustic events, misrecognized speech content, incorrect temporal grounding, unsupported evidence-to-answer binding, or flawed inferences built on audio observations. This creates a need for a benchmark that directly evaluates whether process verifiers can detect erroneous steps, diagnose audio-specific failures, and use process-level signals to improve answer selection.

\begin{table*}[t]
\centering
\scriptsize
\setlength{\tabcolsep}{4pt}
\renewcommand{\arraystretch}{1.25}
\begin{tabular}{lccccccc}
\toprule
\textbf{Benchmark}
& \textbf{PRM}
& \textbf{Audio-}
& \textbf{Step}
& \textbf{Error-Type}
& \textbf{Fine-Grained}
& \textbf{Annotator}
& \textbf{Test Case} \\
& \textbf{Benchmark?}
& \textbf{Grounded?}
& \textbf{Annotation}
& \textbf{Detection?}
& \textbf{Classes}
& 
& \textbf{Size} \\
\midrule




MMAU~\citep{sakshi2025mmau}
& \xmark & \cmark & \xmark & \xmark & -- & Human & 10,000 \\

MMAR~\citep{ma2026mmar}
& \xmark & \cmark & \xmark & \xmark & -- & Human & 1,000 \\

ProcessBench~\citep{zheng2025processbench}
& \cmark & \xmark & \cmark & \xmark & 1 & Human & 3,400 \\

PRMBench~\citep{song2025prmbench}
& \cmark & \xmark & \cmark & \cmark & 9 & Synthetic + Human & 6,216 \\

MPBench~\citep{zhou2025mpbench}
& \cmark & \xmark & \cmark & \xmark & 1 & Synthetic + Human & 9,745 \\

VisualProcessBench~\citep{wang2025visualprm}
& \cmark & \xmark & \cmark & \xmark & 1 & Human & 2,866 \\

\midrule
\textsc{AudioProcessBench}
& \cmark & \cmark & \cmark & \cmark & 6 & Synthetic + Human & 3,872 \\
\bottomrule
\end{tabular}
\caption{
Comparison between \textsc{AudioProcessBench} and related reasoning-process, PRM, multimodal, and audio reasoning benchmarks.
}
\label{tab:benchmark_comparison}
\end{table*}

To address this gap, we introduce \textsc{AudioProcessBench}, a benchmark for process verification in audio-grounded reasoning. \textsc{AudioProcessBench} contains 3,872 semi-automatically annotated reasoning-chain instances collected from 6 audio and omni language model generators. Each trace is segmented into reasoning steps and annotated with step-level correctness labels and fine-grained audio-specific error types. The benchmark supports three complementary evaluation paradigms. First, \textit{step correctness} evaluates whether a verifier can identify erroneous steps in an audio-grounded reasoning chain. Second, \textit{error-type-conditioned detection} analyzes verifier performance across 6 categories of audio reasoning errors: existence, semantic, temporal, acoustic-attribute, cross-modal binding, and reasoning errors. Third, \textit{chain-level aggregation}, inspired by MPBench \citep{zhou2025mpbench}, evaluates whether process-level scores can help select the best final answer from multiple candidate reasoning traces.

We further evaluate 11 audio and omni-modal language models prompted as critic models. Our experiments examine overall verification performance, error-type-specific weaknesses, chain-level aggregation utility, self-critique bias, the relation between generation and criticism, and the effect of in-context examples. The results show that newer and reasoning-oriented models achieve stronger critic performance, but a substantial gap remains between open-source models and the closed-source frontier model. We also find that critic ability is not fully predicted by generation accuracy, and that models may be biased when judging their own reasoning traces.

In summary, our contributions are as follows:
\begin{itemize}
    \item We introduce \textsc{AudioProcessBench}, the first benchmark designed for step-level process verification in audio-grounded reasoning, containing 3,872 reasoning traces and 23,497 annotated steps from 6 audio and omni generators.
    \item We develop an audio-specific process annotation scheme with step-level correctness labels and six fine-grained error types: existence, semantic, temporal, acoustic-attribute, cross-modal binding, and reasoning errors.
    \item We propose three evaluation paradigms for audio process verifiers: step correctness, error-type-conditioned detection with type-conditioned PRMScore, and chain-level answer aggregation over multiple candidate reasoning traces.
    \item We conduct a comprehensive evaluation of 11 critic models, revealing substantial differences across model scale, error types, generator sources, generation-vs-criticism ability, and in-context learning settings.
\end{itemize}

\begin{figure*}[t]
  \centering
  \includegraphics[width=\textwidth]{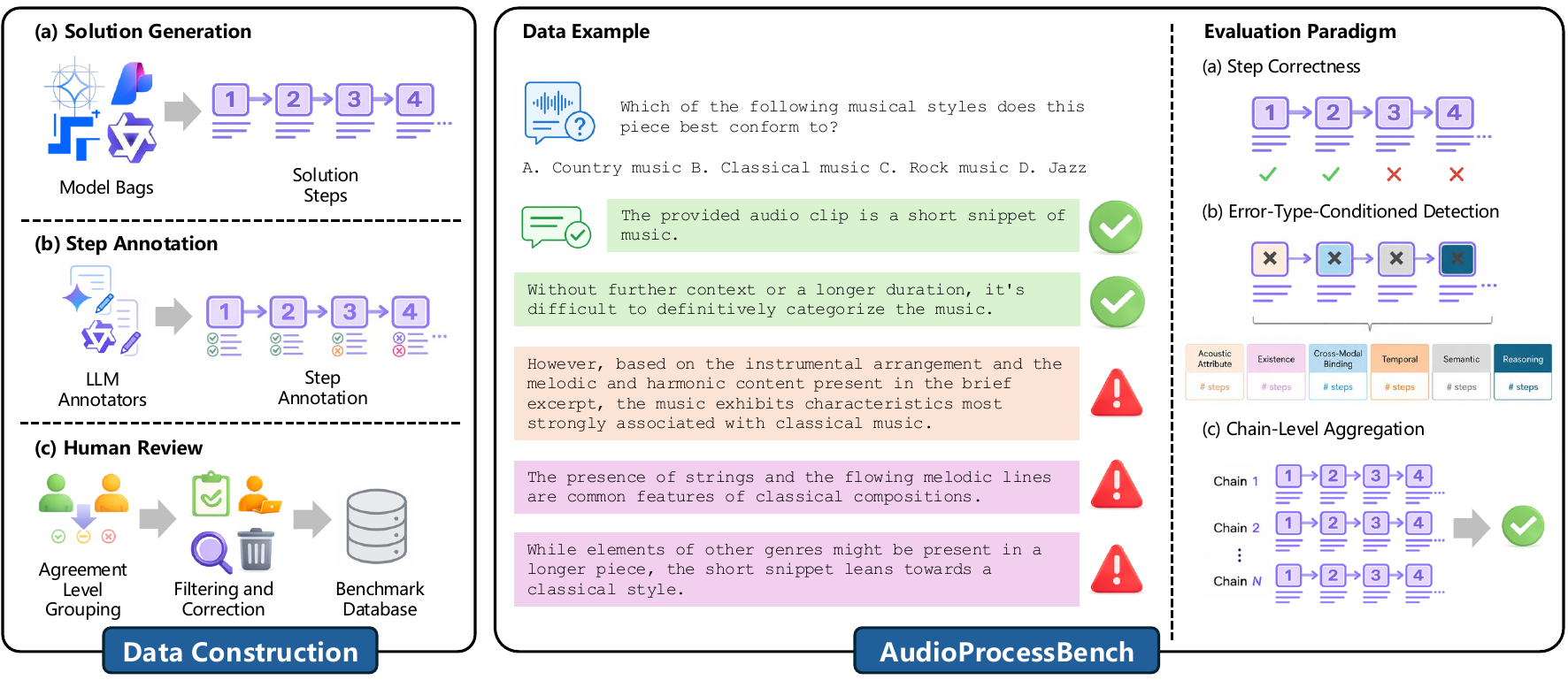}
  \caption{The paradigm of \textsc{AudioProcessBench}. Left: the data construction pipeline of the benchmark; Right: the evaluation paradigm of the benchmark.}
  \label{fig:framework}
\end{figure*}

\section{Related Work}
\subsection{Process-Level Verification in LLM Reasoning}
Verification has become an important mechanism for improving the reliability of LLM reasoning \citep{huang2025pitfalls}. While rule-based verifiers are effective for tasks with executable or formally checkable answers, open-ended reasoning often relies on learned reward models or LLM-as-critic evaluators to assess intermediate steps \citep{lightman2024let}. In particular, process reward models (PRMs) provide step-level feedback rather than only outcome-level supervision, motivating a growing set of process-level verifier benchmarks. Existing benchmarks for PRMs mainly focus on text, mathematics, and vision-language reasoning. ProcessBench evaluates whether models can locate the earliest erroneous step in mathematical reasoning traces \citep{zheng2025processbench}; PRMBench diagnoses PRM capabilities with fine-grained criteria \citep{song2025prmbench}; MPBench and VisualProcessBench extend process-level evaluation to multi-modal and vision-language reasoning \citep{zhou2025mpbench, wang2025visualprm}. However, these benchmarks do not address audio-grounded reasoning, where errors may arise from hallucinated acoustic events, misrecognized speech, incorrect temporal relations, wrong acoustic attributes, or unsupported grounding to the audio signal \citep{sahoo2024comprehensive, cheng2026aha}. As a result, it remains unclear whether existing process verifiers can detect and diagnose errors in audio-based reasoning traces.

\subsection{Audio-Grounded Reasoning and Benchmarks}
Recent audio-language and omni-modal models have shown stronger capabilities to understand speech, music, environmental sounds, and mixed acoustic scenes, and some models can generate explicit reasoning traces for complex audio tasks \citep{chu2024qwen2, abdin2024phi, xu2025qwen2, wu2025step, tian2025step, google_deepmind_gemma3n_2025, google_deepmind_gemma4_2026}. Alongside these models, benchmarks such as MMAU and MMAR evaluate increasingly challenging audio reasoning tasks across diverse audio domains \citep{sakshi2025mmau, ma2026mmar}. Nevertheless, existing audio benchmarks primarily evaluate final-answer accuracy rather than the correctness of intermediate reasoning. They therefore provide limited insight into whether a model's reasoning process is faithful to the acoustic evidence. Moreover, text-only rationales cannot reliably verify audio-specific claims, making it impossible to determine whether a conclusion is properly grounded in the audio. This motivates \textsc{AudioProcessBench}: a step-level benchmark for evaluating audio-grounded reasoning traces, diagnosing audio-specific error types, and testing whether process-level verification can improve final-answer selection among multiple candidate reasoning chains.

\begin{table*}[t]
\centering
\small
\setlength{\tabcolsep}{5pt}
\renewcommand{\arraystretch}{1.12}
\begin{tabular}{lrrrrrrr}
\toprule
& \textbf{Overall} 
& \textbf{Exi.} 
& \textbf{Sem.} 
& \textbf{Temp.} 
& \textbf{Attr.} 
& \textbf{Bind.} 
& \textbf{Rea.} \\
\midrule

\# Steps 
& 23,497 & 3,132 & 1,147 & 526 & 2,600 & 1,455 & 833 \\

Avg. Steps / Trace 
& 6.07 & 0.81 & 0.30 & 0.14 & 0.67 & 0.38 & 0.22 \\

\% of All Steps 
& 100.00 & 13.33 & 4.88 & 2.24 & 11.07 & 6.19 & 3.55 \\

\% of Error Steps 
& 100.00 & 32.31 & 11.83 & 5.43 & 26.82 & 15.01 & 8.59 \\

\bottomrule
\end{tabular}
\caption{
Statistics of the AudioProcessBench. 
Error-type columns correspond to existence error (Exi.), semantic error (Sem.), temporal error (Temp.), acoustic attribute error (Attr.), cross-modal binding error (Bind.), and reasoning error (Rea.).
}
\label{tab:benchmark_statistics}
\end{table*}

\section{Benchmark Construction}
\label{sec:benchmark_construction}

\subsection{Task Definition}
As illustrated in Figure~\ref{fig:framework}, \textsc{AudioProcessBench} evaluates whether a critic model can verify the intermediate reasoning process of an audio-grounded question-answering instance. Given an audio question $\mathbf{Q}$ and a step-by-step reasoning chain $\mathbf{R}=(r_1,r_2,\ldots,r_n)$, the model is required to identify all erroneous steps in the reasoning chain. Formally, the output is a binary judgment array $\mathbf{J}=(j_1,j_2,\ldots,j_n)$, where $j_i \in \{-1,1\}$, with $j_i=-1$ denoting an erroneous step and $j_i=1$ denoting a correct step.

We define an erroneous step as one that contains an objectively incorrect claim, observation, or inference that undermines the validity of the reasoning chain. To support fine-grained analysis, each erroneous step is additionally annotated with an error type. These types are designed to cover both audio-perception failures and reasoning failures, including errors related to event existence, speech semantics, temporal structure, acoustic attributes, cross-modal grounding, and logical inference. The full annotation schema and detailed definitions of all error types are provided in Appendix~\ref{app:annotation_paradigm}.

A key challenge in step-level verification is that later steps may depend on earlier mistakes. \textsc{AudioProcessBench} therefore distinguishes between the original source of an error and its downstream propagation. In general, a later step is labeled as incorrect only when it introduces a new error, adopts a previous error as a factual premise, or actively reinforces an erroneous reasoning path. Otherwise, a step that is locally valid under its immediate context is not automatically penalized simply because an earlier step was wrong. We provide the complete propagation convention in Appendix~\ref{app:annotation_paradigm}.

\subsection{Data Construction}
\subsubsection{Reasoning Generation}
We construct \textsc{AudioProcessBench} from three challenging audio reasoning benchmarks: MMAR \cite{ma2026mmar}, MMSU \citep{dingdong2026mmsu}, and MMAU-Pro \citep{kumar2026mmau}. We select them because they are recent, high-quality, and relatively challenging audio reasoning benchmarks. 
For each selected question, we collect reasoning chains from a diverse set of audio and omni-modal generators. The generator set includes both instruction-following and reasoning-oriented models: Qwen2.5-Omni-7B \citep{xu2025qwen2}, Gemma-3n-E4B \citep{google_deepmind_gemma3n_2025}, Phi-4-Multimodal \citep{abdin2024phi}, Gemma-4-E4B \citep{google_deepmind_gemma4_2026}, Qwen3-Omni-30B-A3B \citep{xu2025qwen3}, and Step-Audio-R1 \citep{tian2025step}. The goal is not to maximize final-answer accuracy, but to obtain diverse reasoning styles and error distributions. The inference generation is conducted using vLLM \citep{kwon2023efficient}. The detailed hyperparameter setting for inference is shown in Appendix~\ref{appendix:infer_param}.

Compared with mathematical reasoning, audio reasoning often has less explicit step boundaries. Therefore, after collecting raw solutions, we use DeepSeek V3.2 \citep{liu2025deepseek} to segment each reasoning chain into discrete reasoning steps. The prompt for step parsing is provided in Appendix~\ref{appendix:parsing_prompt}. After step parsing, we combine rule-based filtering and LLM-based filtering to detect and discard low-quality traces, including repetitive reasoning loops, severe code-switching, obvious output duplication, malformed responses, and chains containing many meaningless or non-informative steps. 

\subsubsection{Step Annotation}
We use two strong closed-source multi-modal annotators, Gemini 3.1 Pro \citep{google_deepmind_gemini31pro_2026} and Qwen3.5 Omni Plus \citep{team2026qwen3}, to annotate step-level correctness and error types. The annotators are instructed to directly label the provided steps and not to redo segmentation. For each step, the annotator outputs a binary correctness label, an error type if the step is incorrect, and a short explanation. The system prompt for the annotators are provided in Appendix \ref{appendix:annotation_prompt}. We choose these models because they are strong audio-capable models and are architecturally heterogeneous. This heterogeneity is important for our annotation pipeline, as we use the agreement between two independently prompted annotators to estimate annotation reliability and route examples into different human review paths. Using two heterogeneous annotators reduces the risk that two similar models make the same systematic judgment error, which could happen if we used two models from the same family.

\subsubsection{Human Review}

After dual-LLM annotation, we divide annotated traces into three groups according to the agreement level between the two LLM annotators: \textit{passed}, \textit{low-disagreement}, and \textit{high-disagreement}. The agreement is computed using the step-level correctness labels. The \textit{passed} group contains traces where the LLM annotators agree on all step correctness labels as well as error types if applicable. The \textit{low-disagreement} group contains traces with minor disagreements within an acceptable range. The \textit{high-disagreement} group contains traces with substantial disagreement, which may indicate difficult reasoning, ambiguous step parsing, or unreliable model annotations. Human annotators perform annotation review and correction after grouping. Details for human annotation are provided in Appendix~\ref{app:annotation_ethics}.

For the \textit{passed} group, we conduct human spot-checking. Specifically, we randomly sample 200 traces and ask two human annotators to independently review them. In this audit set, the two human annotators achieve over 95\% agreement on step-error positions and over 85\% agreement on error types, suggesting that high LLM agreement is a reliable signal for this group. For the \textit{low-disagreement} group, all traces are independently annotated by two human annotators. If the two annotators agree, the trace is retained with the agreed label. If they disagree, a third human annotator adjudicates the case; the trace is retained if at least two annotators agree on the final label, and discarded if all three annotations are inconsistent. The \textit{high-disagreement} group follows the same adjudication strategy, but with an additional pre-review quality inspection. Before human annotation, we first examine the reasoning chain and step parsing quality. Traces with low-quality reasoning, severe parsing errors, or unreliable LLM annotations are discarded. 

\subsection{Evaluation Paradigm}

\textsc{AudioProcessBench} evaluates audio-language and omni-modal models prompted as critic models. Given an audio question and a model-generated reasoning chain, a critic is asked to judge the correctness of each reasoning step. We consider three complementary evaluation paradigms: step correctness, error-type-conditioned detection, and chain-level aggregation.

\paragraph{Step Correctness.}
The primary evaluation measures whether a critic can identify erroneous reasoning steps. For each reasoning chain, the critic outputs a binary correctness judgment for every step, and we additionally evaluate whether it locates the first erroneous step. Following prior process-level benchmarks \citep{zheng2025processbench, song2025prmbench, zhou2025mpbench}, we adopt PRMScore as the main metric for step correctness.

\paragraph{Error-Type-Conditioned Detection.}
This evaluation analyzes whether critics are sensitive to different categories of audio-grounded reasoning errors. Specifically, we do not require critics to predict error types explicitly; they only need to identify erroneous steps. Gold error-type annotations are then used to slice the results by error category, avoiding conflation of step-level error detection with error-type classification, whose boundaries can be subtle. For each of the 6 audio-grounded error types in our taxonomy, we report a type-conditioned PRMScore. This metric evaluates whether a critic can detect a specific type of error while preserving correct steps, without treating other error types as negatives. This evaluation reveals whether a critic's headline performance is uniformly distributed across error categories or instead hides systematic blind spots on specific audio reasoning failures.

\paragraph{Chain-Level Aggregation.}
Following \citet{zhou2025mpbench}, we group multiple reasoning chains generated for the same audio question and evaluate whether critic scores can improve final-answer selection. In the Best-of-$N$ setting, each candidate chain is scored from its step-level judgments, and the final answer of the highest-scoring chain is selected. In majority voting, scores of chains supporting the same answer are aggregated, and the answer with the highest aggregated score is chosen. Both settings are evaluated by final-answer accuracy against the ground truth.

\begin{figure}[t]
  \centering
  \includegraphics[width=\linewidth]{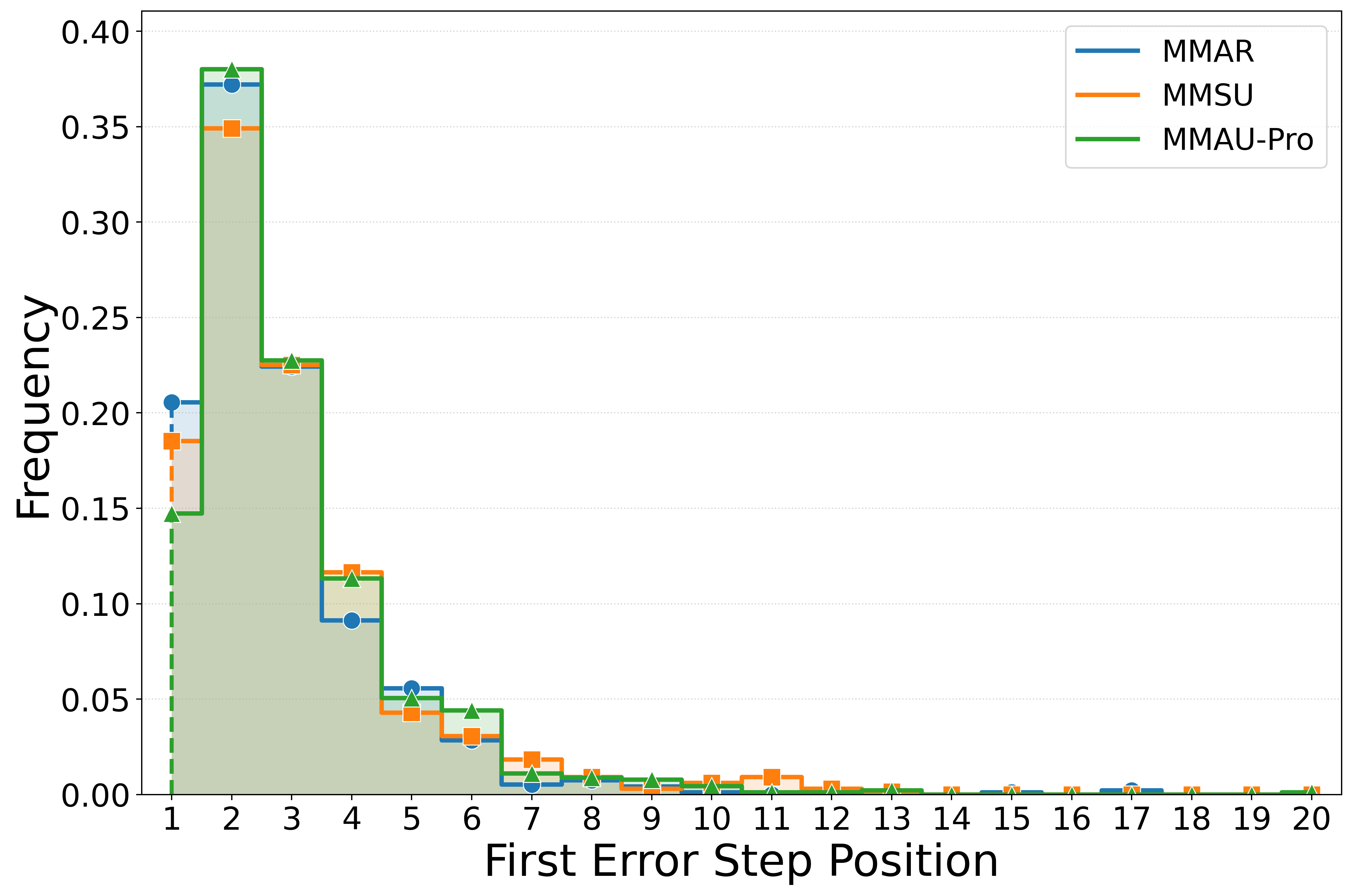}
  \caption{Distribution of the first erroneous step position across source benchmarks.}
  \label{fig:first_error_step_distribution}
\end{figure}

\subsection{Benchmark Statistics}
\label{sec:benchmark_statistics}

\begin{figure}[t]
\centering
\includegraphics[width=\linewidth]{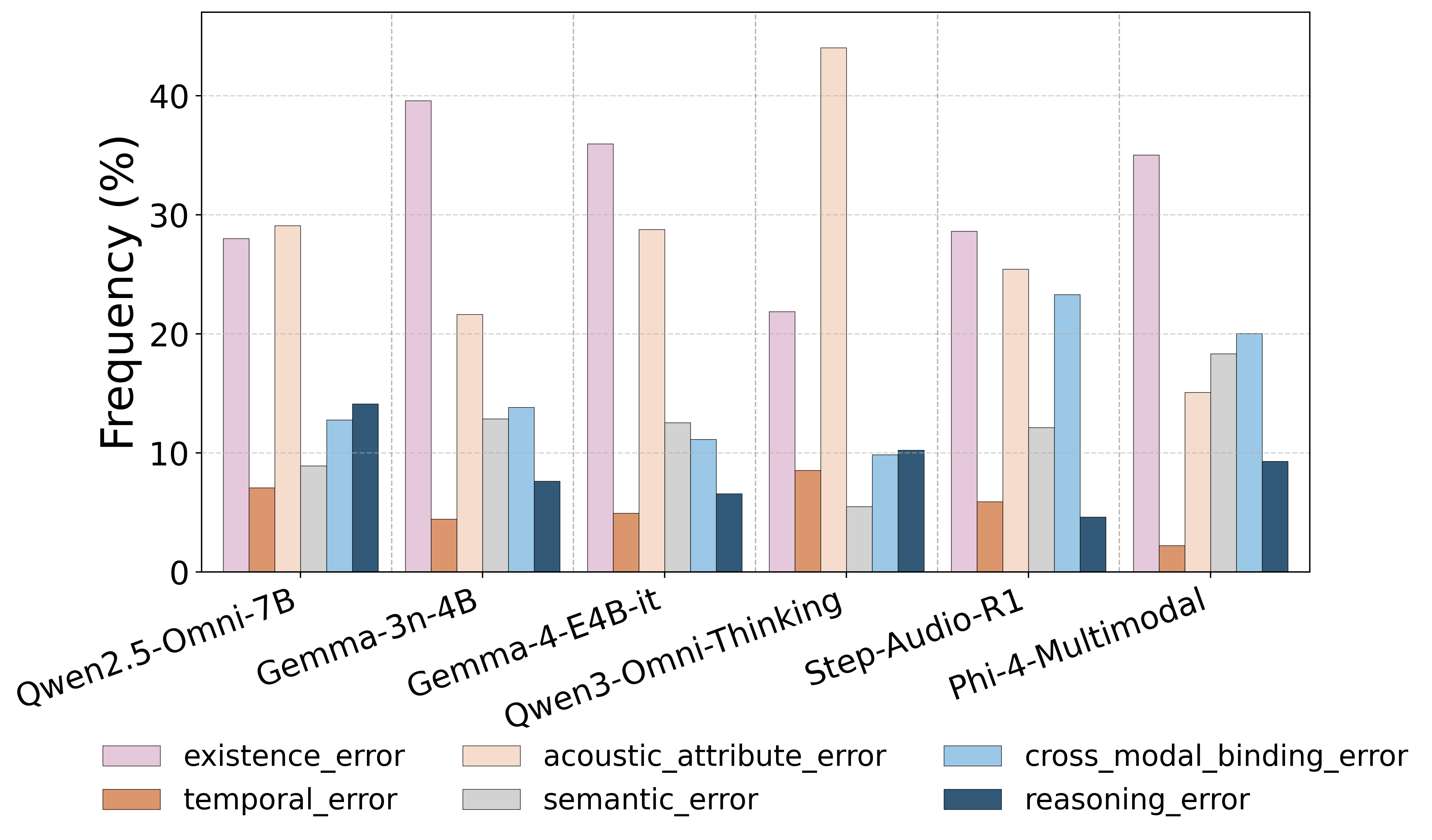}
\caption{Error-type distribution of reasoning traces generated by different models.}
\label{fig:error_type_by_model}
\end{figure}

Table~\ref{tab:benchmark_statistics} summarizes the statistics of \textsc{AudioProcessBench}. The benchmark contains 3,872 reasoning traces and 23,497 annotated steps, with an average trace length of 6.07 steps and a median of 5. The 6 generators achieve 46.64\% final-answer accuracy, indicating that the benchmark is above random guessing but far from saturated. At the step level, 9,693 steps are erroneous, accounting for 41.27\% of all steps. Existence errors are the most frequent category, followed by acoustic-attribute and cross-modal binding errors, suggesting that many audio reasoning failures originate from perception and grounding.

Figure~\ref{fig:first_error_step_distribution} shows that first errors are concentrated in early reasoning steps, especially around steps 2 and 3, with frequency dropping sharply after step 4. This indicates that many reasoning chains become invalid soon after the initial audio observation or grounding stage, motivating step-level verification beyond final-answer evaluation.

\begin{figure}[t]
\centering
\includegraphics[width=\linewidth]{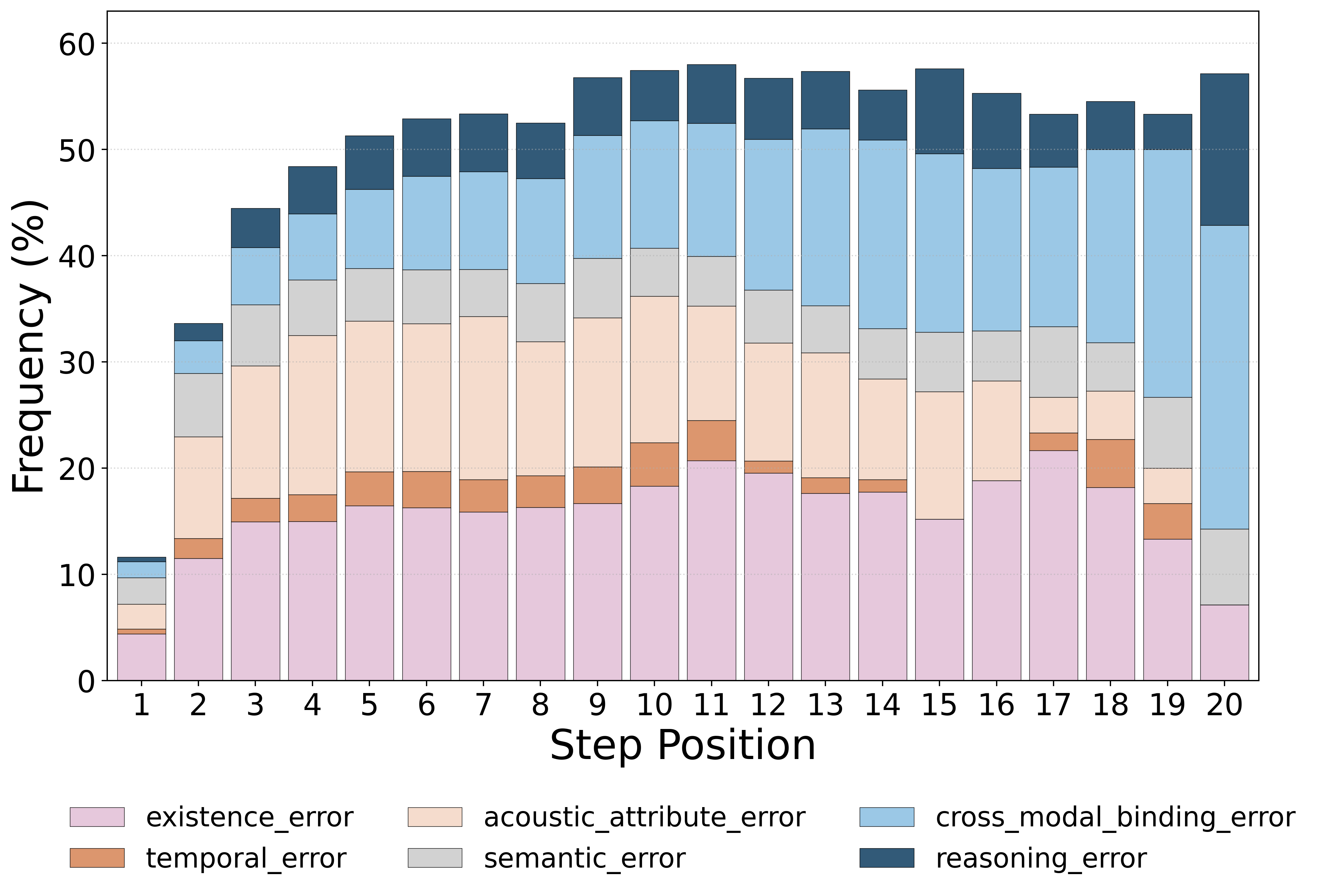}
\caption{Distribution of error types across reasoning step positions.}
\label{fig:error_type_by_position}
\end{figure}

Figures~\ref{fig:error_type_by_model} and~\ref{fig:error_type_by_position} further show that error distributions vary across both generator models and reasoning positions. Different audio and omni language models exhibit distinct failure profiles, while early steps are dominated by existence and acoustic-attribute errors and later steps contain more cross-modal binding and reasoning errors. We do not artificially balance error types, since the imbalance reflects the natural generator-induced failure distribution, and enforcing uniform categories would distort the representativeness of benchmark.

\section{Experiments}
We evaluate a diverse set of audio and omni-modal language models prompted as critic models. All models are prompted with the same critic instruction and are required to judge the correctness of pre-segmented reasoning steps. The system prompt of the critic models is provided in Appendix~\ref{appendix:critic_prompt}. 

\begin{table*}[t]
\centering
\small
\setlength{\tabcolsep}{3pt}
\renewcommand{\arraystretch}{1.12}
\begin{tabular}{l c ccc ccccccc ccc}
\toprule
\multirow{2}{*}{\textbf{Model Name}}
& \multirow{2}{*}{\textbf{Overall}}
& \multicolumn{3}{c}{\textbf{Step Correctness}}
& \multicolumn{7}{c}{\textbf{Error-Type-Conditioned Detection}}
& \multicolumn{3}{c}{\textbf{Chain Aggregation}} \\
\cmidrule(lr){3-5}
\cmidrule(lr){6-12}
\cmidrule(lr){13-15}
&
& \textbf{FEI}
& \textbf{AEI}
& \textbf{Avg.}
& \textbf{Exi.}
& \textbf{Sem.}
& \textbf{Temp.}
& \textbf{Attr.}
& \textbf{Bind.}
& \textbf{Rea.}
& \textbf{Avg.}
& \textbf{BoN}
& \textbf{MV}
& \textbf{Avg.} \\
\midrule
Random
& 37.2
& 2.6 & 44.2 & 23.4
& 44.2 & 43.9 & 44.3 & 44.6 & 44.4 & 43.9 & 44.2
& 47.5 & 40.6 & 44.1 \\
\midrule

\rowcolor{qwenAudioBg}
\href{https://huggingface.co/Qwen/Qwen2-Audio-7B-Instruct}{Qwen2-Audio-7B}
& 34.5\loss{-2.7}
& 8.2 & 27.3 & 17.7
& 34.5 & 36.3 & 34.3 & 31.4 & 44.7 & 28.3 & 34.9
& 50.3 & 51.7 & 51.0 \\

\rowcolor{phiBg}
\href{https://huggingface.co/microsoft/Phi-4-multimodal-instruct}{Phi-4-Multimodal}
& 43.2\gain{6.0}
& 18.5 & 43.5 & 31.0
& 47.7 & 50.3 & 34.1 & 38.4 & 61.9 & 39.0 & 45.2
& 52.9 & 53.7 & 53.3 \\

\rowcolor{qwenOmniBg}
\href{https://huggingface.co/Qwen/Qwen2.5-Omni-3B}{Qwen2.5-Omni-3B}
& 46.4\gain{9.2}
& 25.2 & 49.1 & 37.1
& 51.3 & 44.0 & 43.5 & 43.9 & 59.3 & 48.4 & 48.4
& 52.7 & 54.6 & 53.7 \\

\rowcolor{qwenOmniBg}
\href{https://huggingface.co/Qwen/Qwen2.5-Omni-7B}{Qwen2.5-Omni-7B}
& 42.1\gain{4.9}
& 19.3 & 40.8 & 30.0
& 44.1 & 44.0 & 38.0 & 38.6 & 60.6 & 39.1 & 44.1
& 51.1 & 53.1 & 52.1 \\

\rowcolor{gemma3nBg}
\href{https://huggingface.co/google/gemma-3n-E2B-it}{Gemma-3n-E2B}
& 45.6\gain{8.4}
& 18.8 & 49.2 & 34.0
& 53.6 & 47.6 & 46.3 & 45.2 & 68.1 & 47.7 & 51.4
& 50.3 & 52.5 & 51.4 \\

\rowcolor{gemma3nBg}
\href{https://huggingface.co/google/gemma-3n-E4B-it}{Gemma-3n-E4B}
& 49.7\gain{12.5}
& 25.7 & 55.0 & 40.3
& 56.4 & 56.6 & 47.2 & 50.2 & 72.4 & 55.1 & 56.3
& 50.8 & 54.0 & 52.4 \\

\rowcolor{gemma4Bg}
\href{https://huggingface.co/google/gemma-4-E2B-it}{Gemma-4-E2B}
& 43.6\gain{6.4}
& 24.7 & 44.2 & 34.5
& 44.6 & 38.9 & 35.3 & 33.5 & 64.9 & 43.1 & 43.4
& 52.5 & 53.2 & 52.8 \\

\rowcolor{gemma4Bg}
\href{https://huggingface.co/google/gemma-4-E4B-it}{Gemma-4-E4B}
& 50.5\gain{13.3}
& 35.5 & 54.0 & 44.7
& 51.0 & 55.2 & 52.3 & 42.9 & 73.5 & 55.0 & 55.0
& 51.6 & 52.2 & 51.9 \\

\rowcolor{qwen3Bg}
\href{https://huggingface.co/Qwen/Qwen3-Omni-30B-A3B-Thinking}{Qwen3-Omni-30B-A3B}
& 62.7\gain{25.5}
& \second{54.6} & 68.3 & 61.5
& 69.3 & 68.7 & \second{67.2} & 60.0 & \best{79.4} & 70.2 & 69.2
& \second{57.3} & \second{57.7} & \second{57.5} \\

\rowcolor{stepBg}
\href{https://huggingface.co/stepfun-ai/Step-Audio-R1}{Step-Audio-R1}
& \second{63.2}\gain{26.0}
& 53.7 & \second{71.5} & \second{62.6}
& \second{73.4} & \best{71.4} & \second{67.2} & \second{67.4} & 76.7 & \second{71.8} & \second{71.3}
& 55.0 & 56.2 & 55.6 \\

\rowcolor{geminiBg}
\href{https://ai.google.dev/gemini-api/docs/models/gemini-3-flash-preview}{Gemini-3-Flash}
& \best{67.9}\gain{30.7}
& \best{63.5} & \best{74.8} & \best{69.2}
& \best{76.2} & \second{69.0} & \best{72.1} & \best{71.9} & \second{78.3} & \best{76.8} & \best{74.1}
& \best{59.4} & \best{61.5} & \best{60.5} \\

\bottomrule
\end{tabular}
\caption{
Critic evaluation results on AudioProcessBench.
Rows are lightly shaded by model family.
The green/red subscripts in the \textbf{Overall} column indicate the absolute difference from the random baseline.
The best result in each column is shown in \textbf{bold}, and the second-best result is \underline{underlined}.
\textbf{Step Correctness}: PRMScore for first-error identification (FEI) and all-error identification (AEI).
\textbf{Error-Type-Conditioned Detection}: type-conditioned PRMScore over six audio-grounded error types --- existence (Exi.), semantic (Sem.), temporal (Temp.), acoustic attribute (Attr.), cross-modal binding (Bind.), reasoning (Rea.).
\textbf{Chain-Level Aggregation}: Best-of-$N$ (BoN) and majority-voting (MV) answer-selection accuracy with sum-of-step-scores per trace.
}
\vspace{-2mm}
\label{tab:main_results}
\end{table*}

\subsection{Results}
Table~\ref{tab:main_results} reports the main critic evaluation results on \textsc{AudioProcessBench}. Overall, newer audio and omni models show substantially stronger critic performance than earlier instruction-tuned models, even when the newer models are not always larger in parameter count. For example, the Gemma-3n and Gemma-4 families consistently outperform older audio-oriented models such as Qwen2-Audio-7B and Phi-4-Multimodal, suggesting that more recent multi-modal training recipes improve process-verification ability beyond scale alone. Reasoning-oriented models further strengthen this trend: Qwen3-Omni-30B-A3B and Step-Audio-R1 obtain the best open-source overall scores, with strong gains in both step correctness and error-type-conditioned detection. Nevertheless, a clear gap remains between open-source critics and the closed-source frontier model Gemini-3-Flash, which achieves the highest overall score and leads most evaluation dimensions. This indicates that current open-source models have begun to acquire non-trivial critic ability, but audio-grounded process verification is still far from saturated.

\subsection{Detailed Analysis}
\subsubsection{Model Scale and Critic Performance}
Table~\ref{tab:main_results} shows that larger and newer models generally achieve stronger critic performance, especially among recent audio and omni models. Within the Gemma families, larger variants consistently improve overall verification performance, and the strongest open-source critics are the recent reasoning-oriented models Qwen3-Omni-30B-A3B and Step-Audio-R1, suggesting that both scale and reasoning-oriented training benefit audio-grounded process verification. However, scale alone does not determine critic ability: Qwen2.5-Omni-3B outperforms its 7B variant, and improvements within a family are not uniform across error types. Moreover, gains in step-level verification do not always translate proportionally to chain aggregation, indicating that answer selection also depends on score calibration and candidate-answer distributions.

\begin{figure}[t]
\centering
\includegraphics[width=\linewidth]{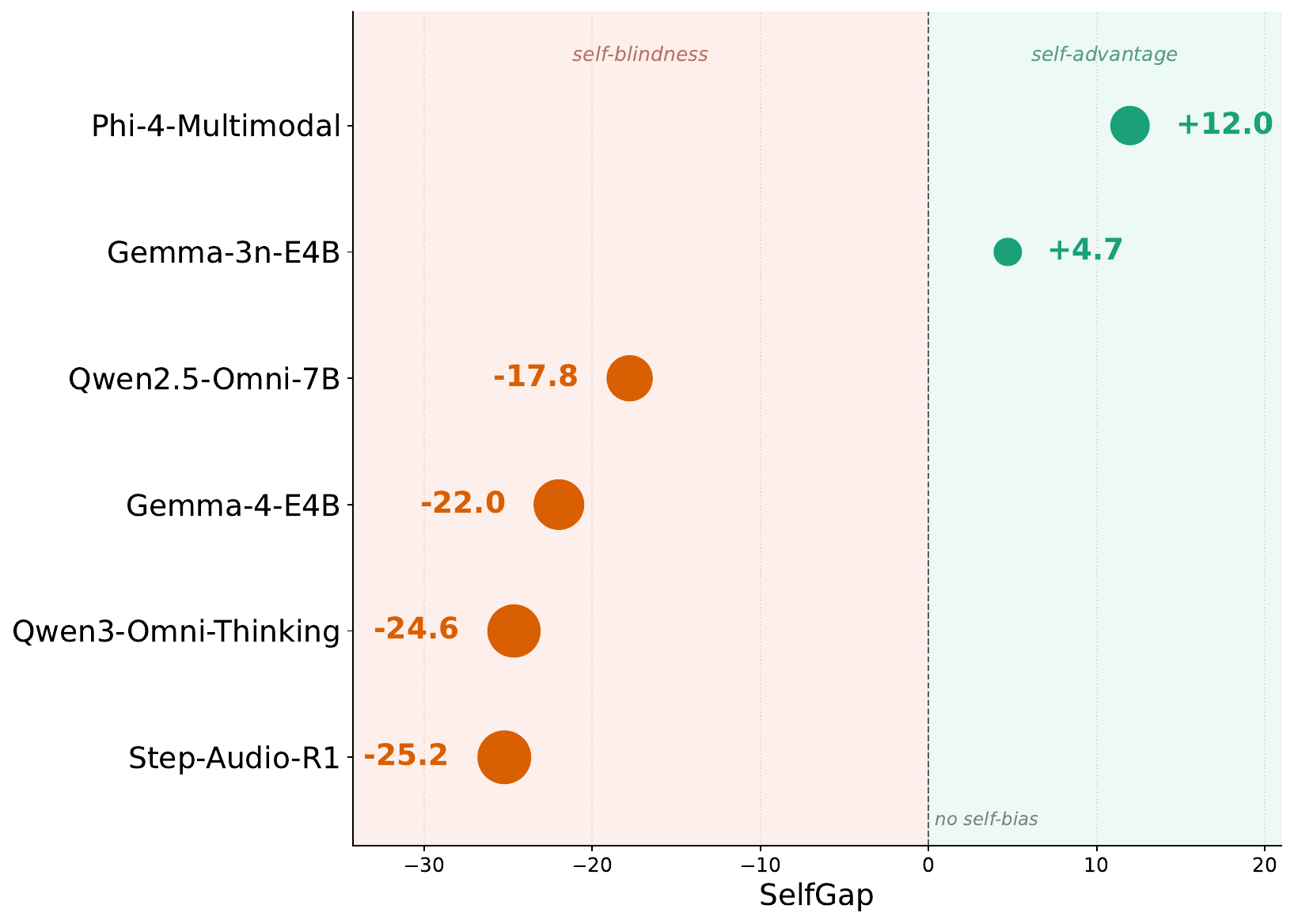}
\caption{Self-critique gap of critic models on \textsc{AudioProcessBench}. 
Positive values indicate self-advantage, while negative values indicate self-blindness. 
The marker size is proportional to the absolute gap magnitude.}
\vspace{-2mm}
\label{fig:self_critique_gap}
\end{figure}

\subsubsection{Error-Type-Specific Critic Performance}
Table~\ref{tab:main_results} shows that critic models exhibit clear type-specific capability profiles. Stronger recent models improve not only on perceptual errors, such as existence and acoustic-attribute errors, but also on higher-level verification abilities reflected in cross-modal binding and reasoning errors, where Qwen3-Omni-30B-A3B, Step-Audio-R1, and Gemini-3-Flash substantially outperform earlier models.

Error-type-specific improvements are also non-uniform across model families. Gemma-3n-E4B and Gemma-4-E4B are already competitive on cross-modal binding errors, while Step-Audio-R1 is particularly strong on semantic, acoustic-attribute, and reasoning errors. Gemini-3-Flash achieves the best average error-type-conditioned performance and remains strong across most categories. These results show that aggregate critic scores can hide important differences in error-type sensitivity, motivating our error-type-conditioned evaluation. In addition, we also report the error-type-specific performance measured by balanced accuracy and AUROC, which is shown in Appendix~\ref{appendix:etcd_appendix}.

\begin{figure}[t]
\centering
\includegraphics[width=\linewidth]{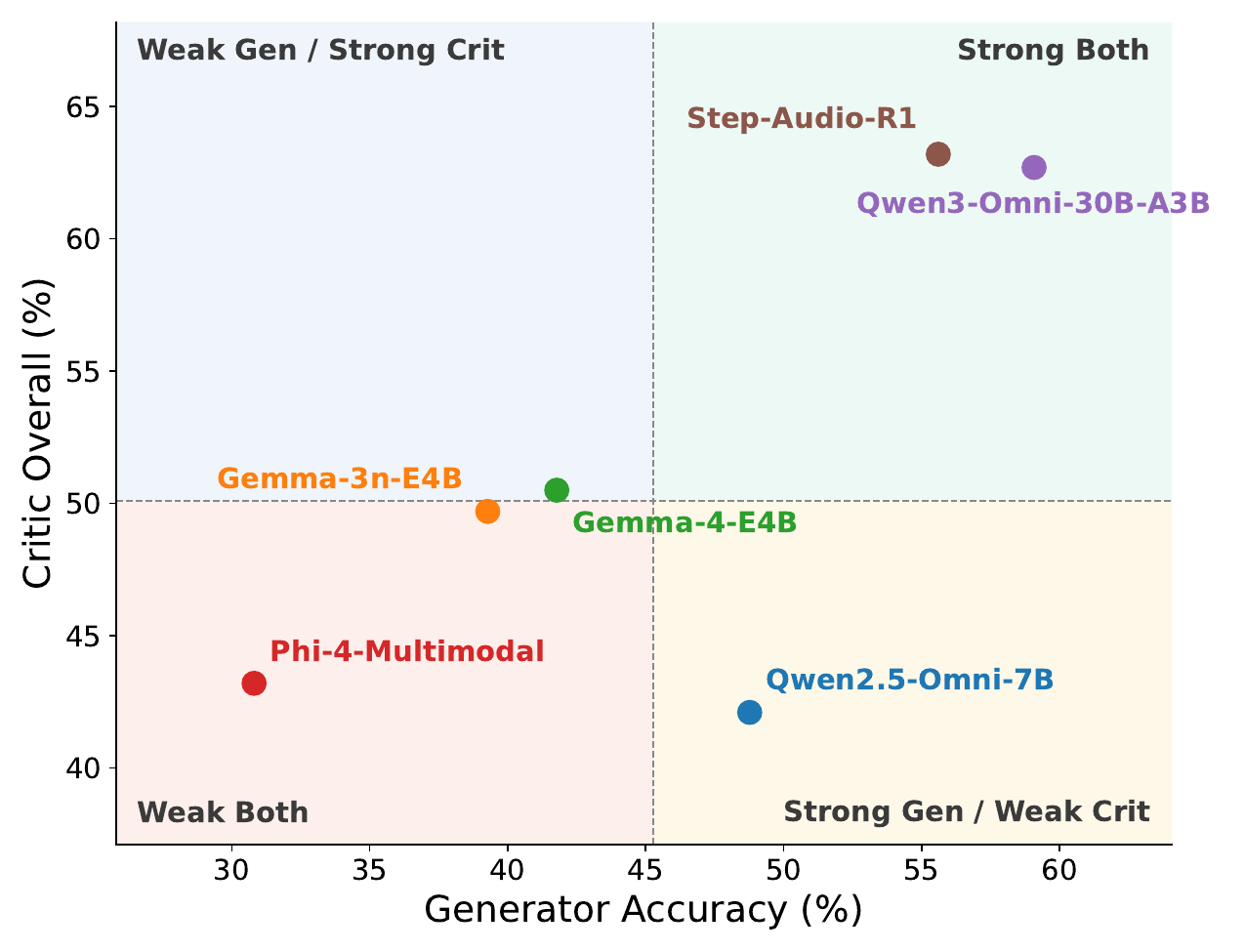}
\caption{Relationship between generation ability and critic ability.}
\label{fig:generation_vs_critic}
\vspace{-2mm}
\end{figure}

\subsubsection{Self-Critique Bias}
We further examine whether critics behave differently when judging their own generated reasoning traces. We define the self-critique gap as the gap of PRMScore on a model's own traces and other generators. A negative value indicates \textit{self-blindness}, where the critic performs worse on its own traces than on traces from other generators. A positive value indicates a \textit{self-advantage}, where the critic is better at judging its own generations.

Figure~\ref{fig:self_critique_gap} shows a clear asymmetry among critic models. Only Phi-4-Multimodal and Gemma-3n-E4B show positive self-critique gaps, suggesting that these models may benefit from familiarity with their own reasoning style or output structure. In contrast, most stronger reasoning-oriented critics exhibit negative gaps. Qwen2.5-Omni-7B, Gemma-4-E4B, Qwen3-Omni-30B-A3B, and Step-Audio-R1 all perform worse on their own generated traces, with the strongest negative gaps observed for Qwen3-Omni-30B-A3B and Step-Audio-R1. This finding suggests that stronger critic ability does not necessarily eliminate self-bias. In fact, the models with the strongest overall critic performance can still show substantial self-blindness, indicating that generation and verification may share similar failure modes within the same model. When a model produces an erroneous reasoning trajectory, its critic mode may be less able to reject the same flawed assumptions or evidence usage. This highlights the importance of evaluating critic models on traces from diverse generators rather than relying only on self-generated reasoning traces.

\begin{table}[t]
\centering
\small
\setlength{\tabcolsep}{6pt}
\renewcommand{\arraystretch}{1.12}
\begin{tabular}{l ccc}
\toprule
\textbf{Model Name} & \textbf{FS} & \textbf{ZS} & \textbf{Gap} \\
\midrule
Random              & 37.2 & 37.2 &  0.0 \\
\midrule
\href{https://huggingface.co/Qwen/Qwen2-Audio-7B-Instruct}{Qwen2-Audio-7B}      & 34.5 & 43.5 & \textcolor{red!70!black}{-9.0} \\
\href{https://huggingface.co/microsoft/Phi-4-multimodal-instruct}{Phi-4-Multimodal}    & 43.2 & 47.3 & \textcolor{red!70!black}{-4.1} \\
\href{https://huggingface.co/Qwen/Qwen2.5-Omni-3B}{Qwen2.5-Omni-3B}     & 46.4 & 46.5 & \textcolor{red!70!black}{-0.1} \\
\href{https://huggingface.co/Qwen/Qwen2.5-Omni-7B}{Qwen2.5-Omni-7B}     & 42.1 & 41.1 & \textcolor{green!60!black}{+1.0} \\
\href{https://huggingface.co/google/gemma-3n-E2B-it}{Gemma-3n-E2B}       & 45.6 & 44.2 & \textcolor{green!60!black}{+1.4} \\
\href{https://huggingface.co/google/gemma-3n-E4B-it}{Gemma-3n-E4B}       & 49.7 & 44.5 & \textcolor{green!60!black}{+5.2} \\
\href{https://huggingface.co/google/gemma-4-E2B-it}{Gemma-4-E2B}        & 43.6 & 40.3 & \textcolor{green!60!black}{+3.3} \\
\href{https://huggingface.co/google/gemma-4-E4B-it}{Gemma-4-E4B}         & 50.5 & 38.1 & \textcolor{green!60!black}{+12.4} \\
\href{https://huggingface.co/Qwen/Qwen3-Omni-30B-A3B-Thinking}{Qwen3-Omni-30B-A3B}  & 62.7 & 58.5 & \textcolor{green!60!black}{+4.2} \\
\href{https://huggingface.co/stepfun-ai/Step-Audio-R1}{Step-Audio-R1}      & 63.2 & 61.8 & \textcolor{green!60!black}{+1.4} \\
\href{https://ai.google.dev/gemini-api/docs/models/gemini-3-flash-preview}{Gemini-3-Flash}     & 67.9 & 64.3 & \textcolor{green!60!black}{+3.6} \\
\bottomrule
\end{tabular}
\caption{
Few-shot vs zero-shot \textbf{Overall} score on AudioProcessBench.
Gap $=$ FS $-$ ZS; positive values indicate that few-shot exemplars help, while negative values indicate the critic performs better without exemplars.
}
\label{tab:fs_vs_zs_overall}
\end{table}
\subsubsection{Generation Ability vs. Critic Ability}
Figure~\ref{fig:generation_vs_critic} compares each model's answer-generation accuracy with its critic performance. Overall, the two abilities are positively related but not equivalent. Qwen3-Omni-30B-A3B and Step-Audio-R1 perform strongly on both axes, suggesting that reasoning-oriented training benefits both answer generation and process verification. However, a mismatch between generation and criticism is also observed. Qwen2.5-Omni-7B achieves relatively strong generation accuracy but much weaker critic performance, while Gemma-3n-E4B and Gemma-4-E4B show competitive critic scores despite more moderate generation accuracy. Phi-4-Multimodal remains weak on both axes. These results indicate that final-answer accuracy alone does not reliably predict critic capability, and that audio-grounded generation and process verification should be evaluated separately.

\subsubsection{Effect of In-Context Learning}
Table~\ref{tab:fs_vs_zs_overall} compares critic performance under few-shot and zero-shot prompting. Few-shot exemplars generally benefit stronger and more recent models, with the largest gain observed for Gemma-4-E4B (+12.4), followed by Gemma-3n-E4B and Qwen3-Omni-30B-A3B. This suggests that stronger critics can better use in-context examples to calibrate their judgments to the annotation protocol. In contrast, earlier or weaker models such as Qwen2-Audio-7B and Phi-4-Multimodal perform better in the zero-shot setting, indicating that few-shot examples may introduce additional instruction-following burden or distract weaker critics from the core verification task. Overall, in-context learning improves critic performance mainly when the model has sufficient capacity to interpret and apply the demonstrations. The full zero-shot results are reported in Appendix~\ref{appendix:critic_zero_shot}.

\section{Conclusion}
We introduce \textsc{AudioProcessBench}, a benchmark for process-level verification in audio-grounded reasoning. It provides step-level correctness labels, audio-specific error annotations, and three complementary evaluation paradigms: step correctness, error-type-conditioned detection, and chain-level aggregation. Evaluating audio and omni-modal models as critics, we find that newer and reasoning-oriented models perform better, though open-source models still lag behind frontier closed-source critics. Critic ability also varies across error types, does not always align with generation ability, and is affected by self-critique bias and in-context learning. These results highlight the need to evaluate audio reasoning beyond final-answer accuracy, and we hope \textsc{AudioProcessBench} can support future work on more reliable audio-grounded process verifiers.

\section*{Limitations}
\textsc{AudioProcessBench} is built on existing multiple-choice audio reasoning benchmarks, and therefore inherits their domain coverage, language distribution, task formats, and data-source biases. While this setup enables controlled verifier evaluation and chain-level aggregation, it does not fully cover open-ended audio reasoning, long-form dialogue, interactive audio understanding, or real-world deployment scenarios. Its current scale is also modest compared with large answer-level audio benchmarks, especially for rare error types.

Our benchmark evaluates explicit model-generated reasoning traces, which may not faithfully reflect models' latent decision processes. Step segmentation and error-type annotation also involve unavoidable ambiguity, since audio observations, semantic interpretation, option comparison, and reasoning are often intertwined. Although we use automatic checks, heterogeneous LLM annotators, agreement-based routing, and human review, residual annotation bias and granularity effects may remain.

Finally, we preserve the natural generator-induced error distribution rather than enforcing class balance. This improves the representativeness of the question-trace pool, but leaves rare categories such as temporal or reasoning errors with fewer examples and higher metric variance. Future work could expand the benchmark scale, cover more diverse audio tasks, and further refine annotation protocols for rare and ambiguous error types.

\section*{Ethics Statement}

\textsc{AudioProcessBench} is constructed from MMAU-Pro \citep{kumar2026mmau}, MMAR \citep{ma2026mmar}, and MMSU \citep{dingdong2026mmsu}. MMAU-Pro and MMAR are released under CC BY-NC 4.0, while MMSU is released under the MIT license. Since part of our benchmark is derived from CC BY-NC 4.0 datasets, \textsc{AudioProcessBench} and its derived annotations will be publicly released under the CC BY-NC 4.0 license for non-commercial research use. We will provide proper attribution to all source benchmarks and document our modifications, including filtering, reasoning generation, step segmentation, and annotation. We will not relicense derived artifacts in a way that removes or weakens the non-commercial restrictions of the source datasets. For audio content originating from third-party platforms or public media, we follow the release format and restrictions of the corresponding source benchmarks and do not claim ownership of the original audio.

\textsc{AudioProcessBench} uses a semi-automatic annotation pipeline with LLM-based segmentation, dual LLM annotation, and human review. To reduce annotation bias and errors, we use heterogeneous LLM annotators, route samples by agreement level, and conduct human auditing or adjudication for uncertain cases. The benchmark is intended for evaluating audio-grounded reasoning process verification, not for surveillance, speaker identification, biometric inference, or privacy-sensitive applications. Potential risks include misuse of the benchmark for evaluating or improving systems in privacy-sensitive audio analysis settings, as well as the propagation of biases inherited from source datasets, selected generator models, LLM annotators, and human reviewers. We therefore recommend using the benchmark only for non-commercial research on reasoning verification and reporting results by source benchmark, task type, generator, and error type when possible.

AI assistants were used for code development, editing, and LaTeX formatting. All AI-assisted outputs were reviewed and revised by the authors, who take full responsibility for the final content, analysis, and claims in this paper.



\bibliography{references}


\appendix

\section{Annotation Paradigm}
\label{app:annotation_paradigm}

\subsection{Error Definition}

In AudioProcessBench, an erroneous step refers to a reasoning step that contains an objectively incorrect claim, observation, or inference that affects the validity of the reasoning chain. A step can be erroneous because it misperceives the audio evidence, misinterprets language-bearing content, incorrectly describes temporal or acoustic properties, grounds the evidence to the wrong target, or performs an invalid inference based on otherwise correct premises.

For each reasoning chain $\mathbf{R}=(r_1,r_2,\ldots,r_n)$, annotators assign a binary judgment to every step. A step is labeled as correct if it is supported by the audio evidence and the preceding valid context, or if it is locally valid and does not introduce or reinforce an error. A step is labeled as incorrect if it contains a new error, adopts an earlier erroneous claim as a factual premise, or further advances an erroneous reasoning path.

\subsection{Error Taxonomy}

Each erroneous step is further annotated with one of the following six error types.

\paragraph{Existence Error.}
An existence error refers to an incorrect claim about whether an audio event, speech event, sound source, speaker, or acoustic cue is present or absent. This includes falsely detecting a sound that is not present, missing a sound that is present, or incorrectly asserting the presence or absence of a relevant acoustic event.

\paragraph{Semantic Error.}
A semantic error refers to an incorrect recognition or interpretation of speech or other language-bearing audio. This includes misheard words, numbers, names, instructions, dialogue content, lyrics, or any other error where the linguistic meaning of the audio is incorrectly represented.

\paragraph{Temporal Error.}
A temporal error refers to an incorrect claim about the temporal structure of the audio. This includes mistakes about event order, timing, duration, simultaneity, repetition, or before/after relations between audio events.

\paragraph{Acoustic Attribute Error.}
An acoustic attribute error refers to an incorrect description of non-semantic acoustic properties. This includes errors about speaker attributes, emotion, tone, instrument, rhythm, pitch, volume, distance, sound quality, or background environment.

\paragraph{Cross-modal Binding Error.}
A cross-modal binding error occurs when a reasoning step makes a decision or intermediate conclusion that is not properly supported by the audio evidence, or when it incorrectly binds audio evidence to the question, answer choice, event role, speaker, or decision target. Unlike semantic or acoustic errors, this category focuses on whether the observed audio evidence is assigned to the correct task-relevant target.

\paragraph{Reasoning Error.}
A reasoning error occurs when the audio observations and premises are mostly correct, but the step makes an invalid inference. This includes logical contradictions, invalid elimination, incorrect comparison, unsupported causal conclusions, arithmetic or counting mistakes, and errors in applying the decision rule required by the question.

\subsection{Local Validity and Error Propagation}

A special challenge in step-level annotation is that later steps may depend on earlier erroneous steps. We therefore distinguish local validity from error propagation.

A step is considered \textit{locally valid} if, given its immediate preceding context, it follows coherently and does not introduce any new unsupported observation or invalid inference. If a locally valid step merely restates or organizes the current reasoning state without further strengthening an erroneous conclusion, we label it as correct, even if an earlier step in the chain was incorrect.

In contrast, a later step is labeled as incorrect if it adopts a previous error as a factual premise and uses it to advance the reasoning chain, eliminate an answer choice, support an intermediate conclusion, or produce an incorrect final decision. In such cases, the propagated step is not automatically labeled as a reasoning error. Instead, its error type inherits the root cause of the original error whenever possible. For example, if an earlier step falsely claims that a dog bark is present and a later step uses this claim to select an answer, the later step is labeled as an existence-related propagated error rather than a pure reasoning error.

This convention allows the benchmark to preserve the distinction between the first source of error and downstream propagation, while still penalizing steps that actively reinforce or operationalize an incorrect conclusion.
\section{Inference Hyperparameters for Solution Generation}
\label{appendix:infer_param}
For inference, all evaluated models are decoded with nucleus sampling probability $p=0.9$ and temperature $t=0.7$. We serve the models using vLLM in a Docker environment based on the \texttt{vllm-cu129-nightly} image. All experiments are run on a Linux server with four NVIDIA H200 141GB GPUs, and all models are uniformly inferred in \texttt{bfloat16} precision.
\section{System Prompt for Solution Step Parsing}
\label{appendix:parsing_prompt}
The steps of the original solutions generated by audio or omni models are parsed using the system prompt shown in Fig. \ref{fig:trace_parsing_prompt}.

\begin{figure*}[t]
\centering
\begin{minipage}{0.96\linewidth}

\begin{promptbox}
{\ttfamily\small
You are given a reasoning trace from an audio language model. Segment it into discrete reasoning steps.

\vspace{0.8em}

CRITICAL RULES:

\vspace{0.6em}

1. PRESERVE THE ORIGINAL TEXT EXACTLY. Do NOT paraphrase, summarize, translate, reorder, or insert any word, character, or punctuation. Each output step must be a verbatim contiguous substring of the input reasoning trace.

\vspace{0.6em}

2. The concatenation of all steps (joined by single space or newline) must reconstruct the original reasoning trace (allowing only inter-step whitespace).

\vspace{0.6em}

3. Step boundaries are decided by natural reasoning flow: one perceptual claim, one inference, one knowledge lookup, one conclusion, etc.

\vspace{0.6em}

4. Don't over-split (every step does real work) or under-split (a step spans multiple distinct reasoning moves).

\vspace{0.8em}

GRANULARITY GUIDANCE:

\vspace{0.6em}

- Aim for 5--12 steps for most reasoning traces; up to 20 only for genuinely very long reasoning (>4000 chars).

\vspace{0.6em}

- Numbered bullets (1., 2., 3.) and bullet markers (-, *) are NOT automatic step boundaries --- group consecutive bullets that belong to the same reasoning phase (e.g. listing observations from a single audio scan, enumerating choices in one analysis pass) into a single step.

\vspace{0.6em}

- Avoid steps shorter than \textasciitilde{}80 characters unless the reasoning move is genuinely atomic (e.g. a one-sentence final inference).

\vspace{0.8em}

Output ONLY JSON (no prose, no fence):

\vspace{0.6em}

\{"steps": ["<step 1 verbatim>", "<step 2 verbatim>", ...]\}
}
\end{promptbox}

\caption{
System prompt for segmenting a solution into discrete steps.
}
\label{fig:trace_parsing_prompt}

\end{minipage}
\end{figure*}
\section{System Prompt for LLM Annotation}
\label{appendix:annotation_prompt}
We use two closed-source flagship omni-modal large language models to perform the annotation of the reasoning steps. The system prompt of the annotators are shown in Fig. \ref{fig:audio_verifier_annotation_prompt_p1} and Fig. \ref{fig:audio_verifier_annotation_prompt_p2}.

\begin{figure*}[t]
\centering
\begin{minipage}{0.96\linewidth}

\begin{promptbox}
{\ttfamily\small
You are an expert audio reasoning verifier. You will receive an audio clip, a question about it, the ground-truth answer, the solver's reasoning segmented into discrete steps, and the solver's final answer. Your job is to (1) label each pre-segmented step as `correct' or as one of six error types, (2) write a 2--4 sentence per-step analysis justifying the label, and (3) report the first-error position and final-answer correctness. Your judgments will become labels in a benchmark for audio reasoning verifier models, so they must be careful, evidence-grounded, and consistent with the protocol below.

\vspace{0.5em}

\textbf{\# 1. Task definition}

\vspace{0.3em}

For each step you receive, in the order given by `step\_id', you will:

\vspace{0.3em}

1. Label the step as exactly one of: `correct', `existence\_error', `temporal\_error', `acoustic\_attribute\_error', `semantic\_error', `cross\_modal\_binding\_error', `reasoning\_error'.

2. Justify the label in a 2--4 sentence `analysis' citing specific audio evidence or logical structure.

3. After labeling all steps, populate `first\_error\_step' and `final\_answer\_correct'.

\vspace{0.3em}

You must label every step exactly once, in the order provided. Do not skip, merge, split, re-order, or invent steps. The set of `step\_id' values in your output must equal the set of `step\_id' values in \textless solver\_segmented\_steps\textgreater{} -- no more, no less.

\vspace{0.5em}

\textbf{\# 2. Step-correctness criterion (load-bearing rule)}

\vspace{0.3em}

A step is labeled as an error if and only if both of the following hold:

\vspace{0.3em}

(a) The step contains an objectively incorrect claim -- perceptual, logical, arithmetic, or a decision unsupported by the audio.

(b) That incorrect claim is either explicitly relied upon by a later step in the chain that derives the final answer, or appears in the final answer itself.

\vspace{0.3em}

If (a) holds but (b) does not -- e.g., the solver makes an incorrect aside that the chain never uses -- label the step `correct'. We are interested in errors that causally affect the trajectory, not incidental misstatements. This matches the ProcessBench convention.

\vspace{0.5em}

\textbf{\# 3. Input format}

\vspace{0.3em}

\textless audio\textgreater{}

The audio clip being reasoned about. 

\textless/audio\textgreater{}

\vspace{0.3em}

\textless question\textgreater{}

The natural-language question posed to the solver.

\textless/question\textgreater{}

\vspace{0.3em}

\textless ground\_truth\_answer\textgreater{}

The verified correct answer to the question. 

\textless/ground\_truth\_answer\textgreater{}

\vspace{0.3em}

\textless solver\_segmented\_steps\textgreater{}

A JSON array of \{``step\_id'': \textless int\textgreater, ``step\_text'': ``\textless verbatim text of the step\textgreater''\} objects, in the order the solver produced them. This segmentation is fixed and authoritative -- do not alter it.

\textless/solver\_segmented\_steps\textgreater{}

\vspace{0.3em}

\textless solver\_final\_answer\textgreater{}

The solver's final answer string. Compare semantically to `ground\_truth\_answer' to set `final\_answer\_correct'.

\textless/solver\_final\_answer\textgreater{}

\vspace{0.5em}

\textbf{\# 4. Error type definitions}

\vspace{0.3em}

\textbf{existence\_error}: The step makes an incorrect claim about whether an audio event, sound source, speech content, speaker, or acoustic cue exists or does not exist in the audio. This includes hallucinating nonexistent audio evidence, missing salient audio evidence, or denying evidence that is actually present. This category is about presence/absence errors.

\vspace{0.3em}

\textbf{temporal\_error}: The step makes an incorrect claim about the temporal structure of the audio or an incorrect derived/computed claim about time, including timestamps, durations, intervals, order, simultaneity, or before/after relations. Time-related errors take precedence over `reasoning\_error' even when the mistake is arithmetic.

\vspace{0.3em}

\textbf{acoustic\_attribute\_error}: The step makes an incorrect claim about non-semantic perceptual attributes, such as speaker attributes, emotion, tone, instrument, rhythm, pitch, volume, distance, background environment, or other acoustic properties.

\vspace{0.3em}

\textbf{semantic\_error}: The step incorrectly recognizes, paraphrases, or interprets the semantic content of speech or language-bearing audio, including words, numbers, names, instructions, dialogue, lyrics, or spoken intent.

\vspace{0.3em}}
\end{promptbox}

\caption{
Prompt template for step-level audio reasoning verification and error-type annotation (Part 1).
}
\label{fig:audio_verifier_annotation_prompt_p1}

\end{minipage}
\end{figure*}

\begin{figure*}[t]
\centering
\begin{minipage}{0.96\linewidth}

\begin{promptbox}
{\ttfamily\small
\textbf{cross\_modal\_binding\_error}: The step makes a decision, selection, or intermediate conclusion that is not supported by the audio evidence. This applies when the individual audio claims are correct, but the leap from those claims to the conclusion or answer choice is not licensed by the audio content.

\vspace{0.3em}

\textbf{reasoning\_error}: All audio perception and cross-modal binding in the step are correct, but the inferential move from the premises to the conclusion is logically, mathematically, or causally defective. Non-temporal arithmetic errors, invalid comparisons, unsupported causal conclusions, and self-contradictions fall here.

\vspace{0.5em}

\textbf{\# 5. Dependent steps: did the step correct the erroneous trajectory, or reinforce it?}

\vspace{0.3em}

Some steps appear after an erroneous step without introducing a new perceptual, binding, or inferential mistake of their own. Such a step is labeled `correct' only if it does not adopt the prior error as a load-bearing premise, or if it actively corrects, retracts, contradicts, or abandons the prior error.

\vspace{0.3em}

A downstream step is labeled as a propagated error if it adopts a prior erroneous claim as a load-bearing premise and continues along the erroneous path. When a downstream step is erroneous only because it propagates an earlier error, assign the root-cause error type of the earlier error, not automatically `reasoning\_error'. If multiple prior errors converge, choose the lowest-layer root cause in the precedence order.

\vspace{0.5em}

\textbf{\# 6. Meta-constraints}

\vspace{0.3em}

- One label per step. Every step receives exactly one of the seven labels.

- Cover every step exactly once, in the same order and with the same `step\_id' values.

- Do not retro-justify from the ground truth. Per-step labels must be derivable from the audio, question, and step content alone.

- Track dependent-step propagation. If a step reinforces a prior uncorrected error into the final answer, label it with the root-cause error type.

- Use epistemic humility for perceptual labels. Before labeling any perceptual error, re-listen to the relevant span at least once. If still uncertain, prefer `correct'.

- For time-related questions, any analysis involving temporal claims must cite specific timestamps observed from re-listening.

\vspace{0.5em}

\textbf{\# 7. Output format}

\vspace{0.3em}

Output only a single JSON object -- no surrounding prose, no markdown code fence, no explanation outside the JSON. Do not echo `step\_text' back.

\vspace{0.3em}

\{
  ``step\_evaluations'': [
    \{
      ``step\_id'': 1,
      ``analysis'': ``\textless 2--4 sentence justification citing specific audio evidence or logical structure.\textgreater'',
      ``label'': ``correct | existence\_error | temporal\_error | acoustic\_attribute\_error | semantic\_error | cross\_modal\_binding\_error | reasoning\_error''
    \}
  ],
  ``first\_error\_step'': \textless integer step\_id of the first non-correct step, or null if all are correct\textgreater,
  ``final\_answer\_correct'': \textless true | false\textgreater
\}

\vspace{0.5em}

\textbf{\# 8. Worked examples}

\vspace{0.3em}

The annotation protocol includes examples for each non-correct error type, one propagated-error case, one correct case with a non-load-bearing slip, and one time-related arithmetic case labeled as `temporal\_error' rather than `reasoning\_error'. These examples illustrate the precedence rule, the propagation rule, and the load-bearing rule used throughout annotation.

\vspace{0.5em}

\textbf{\# 9. Procedure to follow}

\vspace{0.3em}

1. Listen to the entire audio clip end-to-end at least once.

2. Read the question and ground-truth answer to understand the task framing, without using the ground truth to retro-justify step labels.

3. Read the pre-segmented steps in order to understand the chain's trajectory and endpoints.

4. For each step: re-listen to relevant audio spans, decide whether the step contains an objectively incorrect and load-bearing claim, apply the precedence rule, determine whether the step is a fresh error or propagation, and write the analysis.

5. Populate `first\_error\_step' with the smallest non-correct step id, or null if all steps are correct.

6. Set `final\_answer\_correct' by comparing the solver's effective answer to the ground truth semantically.

7. Output the complete JSON object and nothing else.

\vspace{0.5em}

Begin now.
}
\end{promptbox}
\caption{
Prompt template for step-level audio reasoning verification and error-type annotation (Part 2).
}
\label{fig:audio_verifier_annotation_prompt_p2}

\end{minipage}
\end{figure*}
\section{Human Annotation and Data Review Details}
\label{app:annotation_ethics}

\paragraph{Instructions Given to Annotators.}
Human annotators were instructed to review the audio clip, question, answer choices, ground-truth answer, solver final answer, and pre-segmented reasoning steps. For each step, they assigned a binary correctness label and, if the step was incorrect, one of the six error types defined in Appendix~\ref{app:annotation_paradigm}. Annotators were asked to judge whether each step was supported by the audio evidence and the valid preceding context, and to follow the local-validity and error-propagation rules described above. They were allowed to replay the audio when needed and were instructed not to infer unsupported evidence from the final answer alone.

\paragraph{Recruitment and Payment.}
Human review was conducted by members of the research team with experience in audio-language model evaluation and reasoning annotation. No external crowdworkers were recruited. Since the annotation was performed as part of the authors' research activities, no separate per-task payment was involved.

\paragraph{Data Consent and Source Data.}
\textsc{AudioProcessBench} is derived from existing public audio reasoning benchmarks: MMAU-Pro \citep{kumar2026mmau}, MMAR \citep{ma2026mmar}, and MMSU \citep{dingdong2026mmsu}. MMAU-Pro and MMAR are released under CC BY-NC 4.0, while MMSU is released under the MIT license. We follow the usage terms of the source datasets and provide attribution to the original benchmarks. Our annotations are derived from these benchmark instances and are intended for non-commercial research use where required by the original licenses. We do not claim ownership of the original audio content and follow the redistribution format and restrictions of the corresponding source benchmarks. Annotators were informed that their labels would be used to construct and evaluate an academic benchmark for audio-grounded reasoning verification.

\paragraph{Ethics Review Board Approval.}
This work uses publicly released benchmark data and does not collect new personal or sensitive information from human subjects. Human review was limited to annotating model-generated reasoning traces for academic research. Based on our institutional guidelines, this study did not require formal ethics review. We will follow any additional institutional requirements regarding ethics review or exemption before public release of the benchmark.
\section{Prompt Template for Critic Models}
\label{appendix:critic_prompt}
We evaluate audio or omni-modal language models prompted as critic models on our AudioProcessBench. The system prompt used for critic models is provided as Fig. \ref{fig:critic_prompt}. The examples used for in-context learning are provided in Fig.~\ref{fig:fewshot_prompt_ab} and Fig.~\ref{fig:fewshot_prompt_c}.

\begin{figure*}[t]
\centering
\begin{minipage}{0.96\linewidth}

\begin{promptbox}
{\ttfamily\small
You are an expert audio reasoning critic. Your task is to analyze problem-solving steps and provide structured assessments in JSON format.

\vspace{0.8em}

For each step, emit a numerical `score` $\in$ [-1, +1] where:

\vspace{0.6em}

\hspace{1em}- +1 = the step is fully correct, well-grounded in the audio, and advances the chain toward the right answer.

\vspace{0.6em}

\hspace{1em}- 0 = the step is partially correct, ambiguous, or you are unsure (e.g. mixed claims, plausible but unverifiable, harmless meta-commentary that neither helps nor hurts).

\vspace{0.6em}

\hspace{1em}- -1 = the step is clearly wrong (an objective audio / linguistic / inferential mistake that is load-bearing for the chain or is a propagated error from a prior step).

\vspace{0.6em}

\hspace{1em}- Any value in between is allowed; use the full continuous range to express your confidence and the severity of the error.

\vspace{0.8em}

A step is wrong (and deserves a negative score) if and only if both of the following hold:

\vspace{0.6em}

(a) The step contains an objectively incorrect claim --- perceptual, logical, arithmetic, or a decision unsupported by the audio.

\vspace{0.6em}

(b) That incorrect claim is either:

\vspace{0.4em}

\hspace{1em}- explicitly relied upon by a later step in the chain that derives the final answer, OR

\vspace{0.4em}

\hspace{1em}- appears in the final answer itself.

\vspace{0.8em}

Some steps appear after an erroneous step without introducing a new perceptual / binding / inferential mistake of their own. For such a step, the deciding criterion is whether it bends the trajectory back toward the correct answer or continues along the erroneous path.

\vspace{0.8em}

When you assign a negative score for propagation, the magnitude should be comparable to (or slightly smaller than) the root-cause error's score: a step that merely propagates is somewhat less bad than the step that originated the mistake, but still bad. Pick scores like -0.6 to -0.8 for propagated steps when the root cause was scored around -0.8 to -1.0.

\vspace{0.8em}

You will receive:

\vspace{0.6em}

\textless audio\textgreater \\
The audio clip being reasoned about. \\
\textless/audio\textgreater

\vspace{0.6em}

\textless question\textgreater \\
The natural-language question posed to the solver. \\
\textless/question\textgreater

\vspace{0.6em}

\textless solver\_segmented\_steps\textgreater \\
A JSON array of `\{"step\_id": \textless int\textgreater, "step\_text": "\textless verbatim text of the step\textgreater"\}` objects. \\
\textless/solver\_segmented\_steps\textgreater

\vspace{0.6em}

\textless solver\_final\_answer\textgreater \\
The solver's final answer string. \\
\textless/solver\_final\_answer\textgreater

\vspace{0.8em}

Output only a single JSON object --- no surrounding prose, no markdown code fence, no `\textless think\textgreater` block, no explanation outside the JSON. Schema:

\vspace{0.6em}

\{
\quad "step\_evaluations": [
\quad\quad \{"step\_id": 1, "score": 0.85, "analysis": "\textless optional 1--3 sentence justification\textgreater"\},
\quad\quad \{"step\_id": 2, "score": -0.7, "analysis": "..."\}
\quad ]
\}
}
\end{promptbox}

\caption{
System prompt for critic models.
}
\label{fig:critic_prompt}

\end{minipage}
\end{figure*}


\begin{figure*}[t]
\centering
\begin{minipage}{0.96\linewidth}

\begin{promptbox}
{\ttfamily\small

\textbf{Example A --- all-correct solution}

\smallskip

\textbf{<question>}\par
What is the occupation of the first male voice speaking in the video?\par
Choices: A. Firefighter \quad B. Police officer \quad C. Emergency doctor \quad D. Security consultant\par
\textbf{</question>}

\smallskip

\textbf{<solver\_segmented\_steps>}\par
\noindent[
\par
\quad \{``step\_id'': 1, ``step\_text'': ``The first male voice in the video is giving commands and instructions that are typically given by a police officer during an arrest.''\},
\par
\quad \{``step\_id'': 2, ``step\_text'': ``He tells the person to come up, put their hands up, and get on the ground.''\},
\par
\quad \{``step\_id'': 3, ``step\_text'': ``He also uses terms like `hands up' and `get on the ground,' which are standard police procedures.''\},
\par
\quad \{``step\_id'': 4, ``step\_text'': ``Therefore, the occupation of the first male voice speaking in the video is most likely a police officer.''\}
\par
\noindent]
\par
\textbf{</solver\_segmented\_steps>}

\smallskip

\textbf{<solver\_final\_answer>}\par
B. Police officer\par
\textbf{</solver\_final\_answer>}

\smallskip

\textbf{Output:}\par
\noindent\{
\par
\quad ``step\_evaluations'': [
\par
\quad\quad \{``step\_id'': 1, ``score'': 0.95, ``analysis'': ``Correctly characterizes the first male voice as issuing authoritative commands consistent with a law-enforcement arrest.''\},
\par
\quad\quad \{``step\_id'': 2, ``score'': 0.90, ``analysis'': ``Faithful paraphrase of the spoken commands heard in the clip, such as `hands up' and `get on the ground'.''\},
\par
\quad\quad \{``step\_id'': 3, ``score'': 0.95, ``analysis'': ``Correctly links the recognized commands to standard police procedure.''\},
\par
\quad\quad \{``step\_id'': 4, ``score'': 0.90, ``analysis'': ``The conclusion follows from the previous steps and matches the ground-truth answer.''\}
\par
\quad ]
\par
\noindent\}

\medskip

\textbf{Example B --- early error at step 2, propagated}

\smallskip

\textbf{<question>}\par
How many speakers appeared in total in this video?\par
Choices: A. Five \quad B. Two \quad C. Three \quad D. Four\par
\textbf{</question>}

\smallskip

\textbf{<solver\_segmented\_steps>}\par
\noindent[
\par
\quad \{``step\_id'': 1, ``step\_text'': ``Let's analyze the dialogue in the video:''\},
\par
\quad \{``step\_id'': 2, ``step\_text'': ``Speaker 1 and Speaker 2 are listed, omitting the third voice that is plainly audible in the clip.''\},
\par
\quad \{``step\_id'': 3, ``step\_text'': ``We can identify two distinct speakers based on the different lines of dialogue.''\},
\par
\quad \{``step\_id'': 4, ``step\_text'': ``Therefore, there are a total of two speakers in the video.''\}
\par
\noindent]
\par
\textbf{</solver\_segmented\_steps>}

\smallskip

\textbf{<solver\_final\_answer>}\par
B. Two\par
\textbf{</solver\_final\_answer>}

\smallskip

\textbf{Output:}\par
\noindent\{
\par
\quad ``step\_evaluations'': [
\par
\quad\quad \{``step\_id'': 1, ``score'': 0.90, ``analysis'': ``Neutral meta-comment announcing the analysis plan. Harmless and not load-bearing.''\},
\par
\quad\quad \{``step\_id'': 2, ``score'': -0.80, ``analysis'': ``Lists only two speakers and misses a third voice that is clearly present. This is a load-bearing existence error.''\},
\par
\quad\quad \{``step\_id'': 3, ``score'': -0.70, ``analysis'': ``Adopts the undercount from step 2 and reinforces the erroneous two-speaker trajectory.''\},
\par
\quad\quad \{``step\_id'': 4, ``score'': -0.70, ``analysis'': ``Commits to the wrong count as the final answer, directly propagating the step-2 error.''\}
\par
\quad ]
\par
\noindent\}

}
\end{promptbox}

\end{minipage}
\caption{Few-shot examples used for step-level audio reasoning verification. Example A shows an all-correct reasoning chain, while Example B shows an early error that propagates to the final answer.}
\label{fig:fewshot_prompt_ab}
\end{figure*}

\begin{figure*}[t]
\centering
\begin{minipage}{0.96\linewidth}

\begin{promptbox}
{\ttfamily\small
\textbf{Example C --- late error at step 4}

\smallskip

\textbf{<question>}\par
What is the intonation of the entire sentence in the audio?\par
Choices: A. Falling Intonation \quad B. Rise-Fall Intonation \quad C. Fall-Rise Intonation \quad D. Rising Intonation\par
\textbf{</question>}

\smallskip

\textbf{<solver\_segmented\_steps>}\par
\noindent[
\par
\quad \{``step\_id'': 1, ``step\_text'': ``Here's a step-by-step reasoning to determine the intonation of the sentence.''\},
\par
\quad \{``step\_id'': 2, ``step\_text'': ``The audio contains the sentence `Where did you find that?'\,''\},
\par
\quad \{``step\_id'': 3, ``step\_text'': ``Pay attention to how the speaker's voice goes up and down throughout the sentence.''\},
\par
\quad \{``step\_id'': 4, ``step\_text'': ``The final word `that' shows a noticeable rise.''\},
\par
\quad \{``step\_id'': 5, ``step\_text'': ``Rise-Fall Intonation matches the observed pattern.''\},
\par
\quad \{``step\_id'': 6, ``step\_text'': ``The combination indicates a rise-fall, so the answer is B.''\}
\par
\noindent]
\par
\textbf{</solver\_segmented\_steps>}

\smallskip

\textbf{<solver\_final\_answer>}\par
B. Rise-Fall Intonation\par
\textbf{</solver\_final\_answer>}

\smallskip

\textbf{Output:}\par
\noindent\{
\par
\quad ``step\_evaluations'': [
\par
\quad\quad \{``step\_id'': 1, ``score'': 0.95, ``analysis'': ``Meta-comment announcing the analysis plan. Harmless and high-quality.''\},
\par
\quad\quad \{``step\_id'': 2, ``score'': 0.90, ``analysis'': ``Correctly recognizes the spoken sentence.''\},
\par
\quad\quad \{``step\_id'': 3, ``score'': 0.90, ``analysis'': ``Reasonable method for an intonation question.''\},
\par
\quad\quad \{``step\_id'': 4, ``score'': -0.70, ``analysis'': ``First load-bearing perceptual mistake: the actual contour is overall falling, while the step claims a final rise.''\},
\par
\quad\quad \{``step\_id'': 5, ``score'': -0.60, ``analysis'': ``Builds on the step-4 acoustic-attribute error and endorses the wrong option.''\},
\par
\quad\quad \{``step\_id'': 6, ``score'': -0.50, ``analysis'': ``Commits to the wrong final answer based on the propagated pitch-contour error.''\}
\par
\quad ]
\par
\noindent\}

}
\end{promptbox}

\end{minipage}
\caption{Few-shot example used for step-level audio reasoning verification. Example C shows a late perceptual error that propagates to answer selection.}
\label{fig:fewshot_prompt_c}
\end{figure*}
\section{Additional Metrics for Error-Type-Conditioned Detection}
\label{appendix:etcd_appendix}
In the main results, we report type-conditioned PRMScore as the primary metric for error-type-conditioned detection, since it jointly measures the recall of each target error type and the specificity on gold correct steps. To further verify that the observed trends are not tied to this specific harmonic formulation, we additionally report two complementary metrics: balanced accuracy and AUROC. Balanced accuracy provides a threshold-based measure that treats type-specific error steps and correct steps symmetrically, while AUROC evaluates threshold-free ranking ability using the critic's continuous error scores. As shown in Tables~\ref{tab:etcd_ba_fewshot} and~\ref{tab:etcd_auroc_fewshot}, the overall trends are consistent with the main results: stronger recent models, especially Qwen3-Omni-30B-A3B, Step-Audio-R1, and Gemini-3-Flash, remain substantially better across most error types, while earlier models show weaker and more uneven type-specific performance. These results support the robustness of our error-type-conditioned analysis.

\begin{table*}[t]
\centering
\small
\setlength{\tabcolsep}{3pt}
\renewcommand{\arraystretch}{1.12}
\begin{tabular}{l ccccccc}
\toprule
\textbf{Model Name}
& \textbf{Exi.}
& \textbf{Sem.}
& \textbf{Temp.}
& \textbf{Attr.}
& \textbf{Bind.}
& \textbf{Rea.}
& \textbf{Avg.} \\
\midrule
Random
& 50.3 & 49.7 & 50.5 & 51.4 & 50.8 & 49.6 & 50.4 \\
\midrule

\rowcolor{qwenAudioBg}
\href{https://huggingface.co/Qwen/Qwen2-Audio-7B-Instruct}{Qwen2-Audio-7B}
& 51.3 & 52.0 & 51.2 & 50.0 & 55.8 & 48.9 & 51.5 \\

\rowcolor{phiBg}
\href{https://huggingface.co/microsoft/Phi-4-multimodal-instruct}{Phi-4-Multimodal}
& 56.2 & 57.5 & 49.9 & 51.7 & 64.6 & 52.0 & 55.3 \\

\rowcolor{qwenOmniBg}
\href{https://huggingface.co/Qwen/Qwen2.5-Omni-3B}{Qwen2.5-Omni-3B}
& 58.0 & 54.2 & 54.0 & 54.1 & 62.8 & 56.4 & 56.6 \\

\rowcolor{qwenOmniBg}
\href{https://huggingface.co/Qwen/Qwen2.5-Omni-7B}{Qwen2.5-Omni-7B}
& 56.2 & 56.2 & 53.5 & 53.8 & 65.1 & 54.0 & 56.5 \\

\rowcolor{gemma3nBg}
\href{https://huggingface.co/google/gemma-3n-E2B-it}{Gemma-3n-E2B}
& 59.8 & 56.6 & 55.9 & 55.4 & 69.4 & 56.6 & 58.9 \\

\rowcolor{gemma3nBg}
\href{https://huggingface.co/google/gemma-3n-E4B-it}{Gemma-3n-E4B}
& 60.2 & 60.3 & 54.8 & 56.5 & 72.5 & 59.4 & 60.6 \\

\rowcolor{gemma4Bg}
\href{https://huggingface.co/google/gemma-4-E2B-it}{Gemma-4-E2B}
& 57.7 & 55.2 & 53.8 & 53.0 & 68.8 & 57.0 & 57.6 \\

\rowcolor{gemma4Bg}
\href{https://huggingface.co/google/gemma-4-E4B-it}{Gemma-4-E4B}
& 58.6 & 60.9 & 59.3 & 54.4 & 73.9 & 60.8 & 61.3 \\

\rowcolor{qwen3Bg}
\href{https://huggingface.co/Qwen/Qwen3-Omni-30B-A3B-Thinking}{Qwen3-Omni-30B-A3B}
& 72.1 & \second{71.7} & \best{70.7} & 66.2 & \best{80.0} & \second{72.8} & \second{72.2} \\

\rowcolor{stepBg}
\href{https://huggingface.co/stepfun-ai/Step-Audio-R1}{Step-Audio-R1}
& \second{73.5} & 71.6 & \second{68.1} & \second{68.3} & 76.8 & 72.0 & 71.7 \\

\rowcolor{geminiBg}
\href{https://ai.google.dev/gemini-api/docs/models/gemini-3-flash-preview}{Gemini-3-Flash}
& \best{77.8} & \best{73.4} & 66.9 & \best{70.7} & \second{79.1} & \best{76.3} & \best{74.0} \\

\bottomrule
\end{tabular}
\caption{
Per-type balanced accuracy over six audio-grounded error types --- existence (Exi.), semantic (Sem.), temporal (Temp.), acoustic attribute (Attr.), cross-modal binding (Bind.), reasoning (Rea.).
The best result in each column is shown in \textbf{bold}, and the second-best result is \underline{underlined}.
}
\label{tab:etcd_ba_fewshot}
\end{table*}
\begin{table*}[t]
\centering
\small
\setlength{\tabcolsep}{3pt}
\renewcommand{\arraystretch}{1.12}
\begin{tabular}{l ccccccc}
\toprule
\textbf{Model Name}
& \textbf{Exi.}
& \textbf{Sem.}
& \textbf{Temp.}
& \textbf{Attr.}
& \textbf{Bind.}
& \textbf{Rea.}
& \textbf{Avg.} \\
\midrule
Random
& 0.496 & 0.503 & 0.501 & 0.508 & 0.496 & 0.499 & 0.501 \\
\midrule

\rowcolor{qwenAudioBg}
\href{https://huggingface.co/Qwen/Qwen2-Audio-7B-Instruct}{Qwen2-Audio-7B}
& 0.507 & 0.515 & 0.516 & 0.502 & 0.585 & 0.471 & 0.516 \\

\rowcolor{phiBg}
\href{https://huggingface.co/microsoft/Phi-4-multimodal-instruct}{Phi-4-Multimodal}
& 0.530 & 0.553 & 0.454 & 0.469 & 0.641 & 0.474 & 0.520 \\

\rowcolor{qwenOmniBg}
\href{https://huggingface.co/Qwen/Qwen2.5-Omni-3B}{Qwen2.5-Omni-3B}
& 0.588 & 0.550 & 0.532 & 0.540 & 0.627 & 0.567 & 0.568 \\

\rowcolor{qwenOmniBg}
\href{https://huggingface.co/Qwen/Qwen2.5-Omni-7B}{Qwen2.5-Omni-7B}
& 0.582 & 0.574 & 0.559 & 0.545 & 0.690 & 0.549 & 0.583 \\

\rowcolor{gemma3nBg}
\href{https://huggingface.co/google/gemma-3n-E2B-it}{Gemma-3n-E2B}
& \second{0.602} & 0.566 & 0.556 & 0.545 & 0.728 & 0.572 & 0.595 \\

\rowcolor{gemma3nBg}
\href{https://huggingface.co/google/gemma-3n-E4B-it}{Gemma-3n-E4B}
& 0.598 & 0.600 & 0.531 & 0.553 & \second{0.755} & 0.602 & \second{0.606} \\

\rowcolor{gemma4Bg}
\href{https://huggingface.co/google/gemma-4-E2B-it}{Gemma-4-E2B}
& 0.573 & 0.564 & 0.533 & 0.536 & 0.699 & 0.562 & 0.578 \\

\rowcolor{gemma4Bg}
\href{https://huggingface.co/google/gemma-4-E4B-it}{Gemma-4-E4B}
& 0.550 & \second{0.603} & 0.556 & 0.495 & \second{0.755} & 0.581 & 0.590 \\

\rowcolor{qwen3Bg}
\href{https://huggingface.co/Qwen/Qwen3-Omni-30B-A3B-Thinking}{Qwen3-Omni-30B-A3B}
& 0.736 & 0.741 & \best{0.712} & 0.665 & \best{0.826} & 0.732 & 0.735 \\

\rowcolor{stepBg}
\href{https://huggingface.co/stepfun-ai/Step-Audio-R1}{Step-Audio-R1}
& \second{0.752} & 0.737 & 0.690 & \second{0.677} & 0.802 & \second{0.741} & 0.733 \\

\rowcolor{geminiBg}
\href{https://ai.google.dev/gemini-api/docs/models/gemini-3-flash-preview}{Gemini-3-Flash}
& \best{0.774} & \best{0.749} & \second{0.701} & \best{0.689} & 0.815 & \best{0.768} & \best{0.749} \\

\bottomrule
\end{tabular}
\caption{
Per-type AUROC over six audio-grounded error types --- existence (Exi.), semantic (Sem.), temporal (Temp.), acoustic attribute (Attr.), cross-modal binding (Bind.), reasoning (Rea.).
The best result in each column is shown in \textbf{bold}, and the second-best result is \underline{underlined}.
}
\label{tab:etcd_auroc_fewshot}
\end{table*}
\section{Zero-Shot Critic Evaluation}
\label{appendix:critic_zero_shot}

Table~\ref{tab:main_results_zeroshot} reports the full critic evaluation results under the zero-shot prompt setting. Compared with the few-shot setting in the main results, zero-shot prompting leads to mixed effects across model families. Earlier or smaller models such as Qwen2-Audio-7B and Phi-4-Multimodal perform better without exemplars, suggesting that few-shot demonstrations may introduce additional instruction-following burden for weaker critics. In contrast, stronger reasoning-oriented or more recent models, including Qwen3-Omni-30B-A3B, Step-Audio-R1, and Gemini-3-Flash, remain highly competitive in zero-shot settings but generally benefit from few-shot prompting in the main evaluation. These results indicate that in-context examples are helpful mainly when the critic has sufficient capacity to interpret and apply the annotation protocol.

\begin{table*}[t]
\centering
\small
\setlength{\tabcolsep}{3pt}
\renewcommand{\arraystretch}{1.12}
\begin{tabular}{l c ccc ccccccc ccc}
\toprule
\multirow{2}{*}{\textbf{Model Name}}
& \multirow{2}{*}{\textbf{Overall}}
& \multicolumn{3}{c}{\textbf{Step Correctness}}
& \multicolumn{7}{c}{\textbf{Error-Type-Conditioned Detection}}
& \multicolumn{3}{c}{\textbf{Chain Aggregation}} \\
\cmidrule(lr){3-5}
\cmidrule(lr){6-12}
\cmidrule(lr){13-15}
&
& \textbf{FEI}
& \textbf{AEI}
& \textbf{Avg.}
& \textbf{Exi.}
& \textbf{Sem.}
& \textbf{Temp.}
& \textbf{Attr.}
& \textbf{Bind.}
& \textbf{Rea.}
& \textbf{Avg.}
& \textbf{BoN}
& \textbf{MV}
& \textbf{Avg.} \\
\midrule
Random
& 37.2
& 2.6 & 44.2 & 23.4
& 44.2 & 43.9 & 44.3 & 44.6 & 44.4 & 43.9 & 44.2
& 47.5 & 40.6 & 44.1 \\
\midrule

\rowcolor{qwenAudioBg}
\href{https://huggingface.co/Qwen/Qwen2-Audio-7B-Instruct}{Qwen2-Audio-7B}
& 43.5\gain{6.3}
& 25.1 & 48.2 & 36.7
& 49.1 & 48.1 & 37.9 & 38.3 & 55.8 & 37.8 & 44.5
& 46.9 & 51.5 & 49.2 \\

\rowcolor{phiBg}
\href{https://huggingface.co/microsoft/Phi-4-multimodal-instruct}{Phi-4-Multimodal}
& 47.3\gain{10.1}
& 26.2 & 52.3 & 39.2
& 55.9 & 51.3 & 45.8 & 44.3 & 66.1 & 42.0 & 50.9
& 50.2 & 53.5 & 51.9 \\

\rowcolor{qwenOmniBg}
\href{https://huggingface.co/Qwen/Qwen2.5-Omni-3B}{Qwen2.5-Omni-3B}
& 46.5\gain{9.3}
& 28.0 & 48.7 & 38.4
& 50.4 & 48.0 & 46.5 & 46.1 & 55.9 & 43.6 & 48.4
& 50.6 & 54.5 & 52.6 \\

\rowcolor{qwenOmniBg}
\href{https://huggingface.co/Qwen/Qwen2.5-Omni-7B}{Qwen2.5-Omni-7B}
& 41.1\gain{3.9}
& 18.5 & 41.0 & 29.8
& 44.1 & 43.8 & 36.7 & 33.6 & 50.2 & 33.6 & 40.3
& 51.7 & 54.9 & 53.3 \\

\rowcolor{gemma3nBg}
\href{https://huggingface.co/google/gemma-3n-E2B-it}{Gemma-3n-E2B}
& 44.2\gain{7.0}
& 22.7 & 46.3 & 34.5
& 49.9 & 41.3 & 36.7 & 37.8 & 64.3 & 42.2 & 45.3
& 51.6 & 54.2 & 52.9 \\

\rowcolor{gemma3nBg}
\href{https://huggingface.co/google/gemma-3n-E4B-it}{Gemma-3n-E4B}
& 44.5\gain{7.3}
& 21.4 & 46.8 & 34.1
& 46.8 & 40.0 & 42.9 & 37.1 & 66.5 & 47.3 & 46.8
& 51.4 & 53.9 & 52.6 \\

\rowcolor{gemma4Bg}
\href{https://huggingface.co/google/gemma-4-E2B-it}{Gemma-4-E2B}
& 40.3\gain{3.1}
& 19.6 & 39.4 & 29.5
& 38.4 & 35.3 & 33.7 & 31.9 & 58.5 & 36.8 & 39.1
& 52.1 & 52.6 & 52.3 \\

\rowcolor{gemma4Bg}
\href{https://huggingface.co/google/gemma-4-E4B-it}{Gemma-4-E4B}
& 38.1\gain{0.9}
& 29.3 & 31.2 & 30.2
& 30.4 & 36.3 & 29.4 & 22.1 & 37.0 & 44.2 & 33.2
& 50.6 & 51.2 & 50.9 \\

\rowcolor{qwen3Bg}
\href{https://huggingface.co/Qwen/Qwen3-Omni-30B-A3B-Thinking}{Qwen3-Omni-30B-A3B}
& 58.5\gain{21.3}
& \second{55.7} & 61.1 & 58.4
& 65.7 & 65.1 & 61.4 & 51.9 & 63.1 & 61.2 & 61.4
& \best{54.1} & \best{57.4} & \best{55.8} \\

\rowcolor{stepBg}
\href{https://huggingface.co/stepfun-ai/Step-Audio-R1}{Step-Audio-R1}
& \second{61.8}\gain{24.6}
& 52.7 & \second{70.4} & \second{61.5}
& \best{72.5} & \best{69.8} & \second{67.5} & \second{65.9} & \second{72.3} & \best{72.9} & \second{70.1}
& \second{52.9} & 54.5 & 53.7 \\

\rowcolor{geminiBg}
\href{https://ai.google.dev/gemini-api/docs/models/gemini-3-flash-preview}{Gemini-3-Flash}
& \best{64.3}\gain{27.1}
& \best{61.2} & \best{72.7} & \best{66.9}
& \second{70.6} & \second{66.3} & \best{71.1} & \best{68.4} & \best{76.5} & \second{71.0} & \best{70.7}
& \best{54.1} & \second{56.5} & \second{55.3} \\

\bottomrule
\end{tabular}
\caption{
Critic evaluation results on AudioProcessBench under zero-shot prompt settings.
The green/red subscripts in the \textbf{Overall} column indicate the absolute difference from the random baseline.
The best result in each column is shown in \textbf{bold}, and the second-best result is \underline{underlined}.
\textbf{Step Correctness}: PRMScore for first-error identification (FEI) and all-error identification (AEI).
\textbf{Error-Type-Conditioned Detection}: type-conditioned PRMScore over six audio-grounded error types --- existence (Exi.), semantic (Sem.), temporal (Temp.), acoustic attribute (Attr.), cross-modal binding (Bind.), reasoning (Rea.).
\textbf{Chain-Level Aggregation}: Best-of-$N$ (BoN) and majority-voting (MV) answer-selection accuracy with sum-of-step-scores per trace.
}
\label{tab:main_results_zeroshot}
\end{table*}

\end{document}